\documentclass[a4paper,11pt]{article}

\pdfoutput=1 

\usepackage{jcappub} 

\title{ \boldmath Structure formation in inhomogeneous Early Dark Energy models}

\author[a]{R. C. Batista}
\author[b]{and F. Pace}

\affiliation[a]{Escola de Ci\^encias e Tecnologia, Universidade Federal do Rio
Grande do Norte\\
Caixa Postal 1524, 59072-970, Natal, Rio Grande do Norte, Brazil}
\affiliation[b]{Institute of Cosmology and Gravitation, University of
Portsmouth,\\ Dennis Sciama Building, Portsmouth, PO1 3FX, U.K.}

\emailAdd{rbatista@ect.ufrn.br}
\emailAdd{francesco.pace@port.ac.uk}

\abstract{
We study the impact of Early Dark Energy fluctuations in the linear and
non-linear regimes of structure formation. In these models the energy density
of dark energy is non-negligible at high redshifts and the fluctuations in the
dark energy component can have the same order of magnitude of dark matter
fluctuations. Since two basic approximations usually taken in the standard
scenario of quintessence models, that both dark energy density during the matter
dominated period and dark energy fluctuations on small scales are negligible,
are not valid in such models, we first study approximate
analytical solutions for dark matter and dark energy perturbations in the linear
regime. This study is helpful to find consistent initial conditions for the
system of equations and to analytically understand the effects of Early Dark
Energy and its fluctuations, which are also verified numerically. In the linear
regime we compute the matter growth and variation of the gravitational potential
associated with the Integrated Sachs-Wolf effect, showing that these observables
present important modifications due to Early Dark Energy fluctuations, though
making them more similar to $\Lambda$CDM model. We also make use of the
Spherical Collapse model to study the influence of Early Dark Energy
fluctuations in the nonlinear regime of structure formation, especially on
$\delta_c$ parameter, and their contribution to the halo mass, which we show can
be of the order of $10\%$. We finally compute how the number density of halos is
modified in comparison to $\Lambda$CDM model and address the problem of how to
correct the mass function in order to take into account the contribution of
clustered dark energy. We conclude that the inhomogeneous Early Dark Energy
models are more similar to $\Lambda$CDM model than its homogeneous counterparts. 
}

\begin{document}
\maketitle
\flushbottom

\section{Introduction}
The understanding of the accelerated expansion of the universe is one of the
greatest challenges in physics. If we assume that General Relativity is our
accepted theory of gravitation we must introduce a new form of fluid with
sufficiently negative pressure, $p<-\rho/3$, that accounts for roughly $3/4$ of
the universe energy density today. The physical description of this new form of
matter, generically called dark energy (DE), is yet unknown. Latest data
analysis of luminosity distance of Supernovae type Ia \cite{Amanullah2010},
cosmic microwave background \cite{Hinshaw:2012fq,Sievers:2013wk} and large scale
structure \cite{Reid2010c,Blake:2011rj} are all consistent with a flat universe
with approximately $1/4$ of its critical density in the form of pressureless
matter (cold dark matter and baryons) and $3/4$ in the Cosmological Constant,
$\Lambda$.

If the accelerated expansion is caused by $\Lambda$, since it is constant in
space and time, it does not cluster and has a negligible contribution to the
energy density budget of the universe at high redshifts, affecting solely the
background evolution for $z \sim 1$ and lower. Although $\Lambda$ is the
simplest model for the accelerated expansion, it suffers from two severe
theoretical problems. Since the natural interpretation of $\Lambda$ is the
vacuum energy, Quantum Field Theory should be able to determine its value.
However it predicts a value for $\Lambda$ that can be
several tens of orders of magnitude larger than what is observed, which is known
as the Cosmological Constant Problem,
see, e.g., \cite{Weinberg1989a,Sahni2000}. Another problem, more closely related
to the cosmological evolution itself,
is why the observed value of $\Lambda$ is such that it becomes important for the
evolution of the universe just at the time we are able to measure its effects,
which is known as Coincidence Problem \cite{Zlatev:1998tr}.

Many alternative models for the accelerated expansion have been proposed.
Possibly the most studied ones are based on
canonical scalar fields \cite{Peebles:1987ek,Ratra:1987rm,Wetterich:1987fm},
which are usually referred as quintessence models. The evolution of quintessence
is background dependent, fact that could explain the transition
from a decelerated to an accelerated phase as a natural evolution between
attractor regimes \cite{Zlatev:1998tr,Steinhardt:1999nw}. This kind of mechanism
can potentially alleviate the Coincidence Problem, diminish the dependence on
the initial conditions of the scalar field and bring new features to the
cosmological evolution, for instance, the possibility
that the energy density of DE is non-negligible at high redshifts. 
Models presenting this behaviour are called Early Dark Energy (EDE) and were
extensively studied in the literature, see for instance
\cite{Doran2006,Bartelmann2006,Francis2008,Calabrese2011a,Reichardt2012}.

Another important feature of dynamical DE models is that, in contrast with
$\Lambda$, they possess fluctuations. On small scales, in the linear regime,
quintessence perturbations are several orders of magnitude smaller than Dark
Matter (DM) perturbations and are usually neglected in studies of structure
formation. This is due to the fact that the effective sound speed of canonical
scalar fields perturbations, or the sound speed in the rest frame
\cite{Hu1998c}, is $c_{{\rm eff}}=\delta p_{\rm e}/\delta\rho_{\rm e}=1$, which
suppresses the growth of field perturbations inside the sound horizon scale,
which, in turn, is of order of the particle horizon. 
However quintessence fluctuations can not be neglected from both the theoretical
and observational point of view
\cite{Caldwell:1997ii,Park2009,Christopherson:2010jz}.

Moreover, there exists some realisations of DE models where its fluctuations can
grow on sub-horizon scales, i.e., with $c_{{\rm eff}}\ll 1$, such as k-essence
models \cite{Chiba2000,Armendariz-Picon2001,Chimento2005c,Creminelli2009}. The
possibility that DE has an effective sound speed less than unity has been
investigated by many authors, e.g.,
\cite{Erickson2002,Bean2004,Hu2004,Sapone2009,Putter2010}. In particular,
Ref.~\cite{Putter2010} points out that current CMB and LLS data slightly prefers
dynamical DE, $c_{\rm eff} \neq 1$ and some amount of EDE. It is also worth to
note that clustered DE seems to give a better prediction for the concentration
parameter of massive galaxy clusters \cite{Basilakos2009c}. 

Structure formation in EDE models has been studied by many authors, e.g.,
\cite{Bartelmann2006,Francis2008,Grossi2009,Xia:2009ys,Pace2010,Alam2010,
Alam2011,Wang:2012fq}. In particular, for the case
of $ c_{{\rm eff}}=1$, Ref. \cite{Alam2010} shows that neglecting EDE
perturbations leads to incorrect constrains on the equation of state and Ref.
\cite{Alam2011} claims that EDE models can be constrained by
future observations of galaxy clusters. However all these
studies either neglected EDE perturbations or consider models with $ c_{{\rm
eff}}=1$, which effectively renders negligible perturbations on small scales.

The objective of this paper is to analyse structure formation in EDE models that
can present large fluctuations on small scales. In this scenario DE has
two major characteristics not present at the same time in the usual quintessence
models: 1) DE energy density is non-negligible at high$-z$ and 2) DE
fluctuations can be of the same order of magnitude of DM fluctuations. We
compare the results with the usual assumption of nearly homogeneous EDE and
$\Lambda$CDM model. Assuming that EDE is described by a
perfect fluid, characterised by its equations of state, $p_e=w(t)\rho_{\rm e}$,
and the effective sound speed of its perturbations $c^2_{\rm eff}=\delta p_{\rm
e}/\delta\rho_{\rm e}$, we analyse both the linear and nonlinear evolution of
EDE fluctuations and their impact on DM growth, compute the number density of
halos and how it is modified by the contribution of DE fluctuations.

The outline of the paper is the following. In Sect.~\ref{sect:back_evol} we
present and discuss the background evolution of two models of EDE. In
Sect.~\ref{sect:lin_evol} we study the evolution of linear perturbations of a
system with EDE and pressureless matter and calculate the matter growth and the
Integrated Sachs-Wolf effect. Sect.~\ref{sect:nl_evol} is devoted to the study
of the nonlinear evolution and the Spherical Collapse Model. In
Sect.~\ref{sect:mf} we present the results for mass functions and in
Sect.~\ref{sect:conc} we present our conclusions.

\section{Background evolution}\label{sect:back_evol}

We assume a universe with flat spatial section, DM (baryons are treated as dark
matter) and DE. Friedmann's equations in conformal time are then:
\begin{equation}
\mathcal{H}^{2}=\frac{8\pi G}{3}a^{2}\left(\rho_{\rm m}+\rho_{\rm
e}\right)\,\,\,\,
\mbox{and}\,\,\,\dot{\mathcal{H} }=-\frac{4\pi G}{3}a^{2}\left[\rho_{\rm
m}+\rho_{\rm e}\left(1+3w\right)\right]\,,
\end{equation}
where $\mathcal{H}=\dot{a}/a$ and the dots represent derivative with respect to
conformal time. We choose two different parametrizations for EDE. In the first
one the energy density parameter of DE, $\Omega_{\rm e}=8\pi G a^{2}\rho_{\rm
e}/3\mathcal{H}^{2}$, is given as a function of its value at early times,
$\Omega_{\rm e}^{\rm e}$, its equation-of-state parameter now, $w_{0}$, and
matter energy density now,
$\Omega_{\rm m}^{0}$, \cite{Doran2006}:
\begin{equation}
\Omega_{\rm e}\left(a\right)=\frac{\Omega_{\rm e}^{0}-
\Omega_{\rm e}^{\rm e}\left(1-a^{-3w_{0}}\right)}{\Omega_{\rm e}^{0}+
\Omega_{\rm m}^{0}a^{3w_{0}}}+\Omega_{\rm e}^{\rm
e}\left(1-a^{-3w_{0}}\right)\,.
\end{equation}
We will refer to this parametrization as Model A. The advantage of this
parametrization is that one can fix
$\Omega_{\rm e}^{\rm e}$, however it does not allow to directly choose
details of evolution of the equation-of-state
parameter. In order to control some properties of $w$, such as the value in the
matter dominated era, $w_{\rm m}$, the
moment of transition from $w_{\rm m}$ to $w_{0}$, $a_{\rm c}$, and the duration
of this transition, $\Delta_{\rm m}$, we
also study the following parametrization \cite{Corasaniti2003}:
\begin{equation}
w\left(a\right)=w_{0}+\left(w_{\rm
m}-w_{0}\right)\frac{1+\exp{\left(\frac{a_{\rm c}}{\Delta_{\rm m}}\right)}}
{1+\exp{\left(-\frac{a-a_{\rm c}}{\Delta_{\rm m}}\right)}}
\frac{1-\exp{\left(\frac{a-1}{\Delta_{\rm
m}}\right)}}{1-\exp{\left(\frac{1}{\Delta_{\rm m}}\right)}}\,,
\label{eq:corass_parm}
\end{equation}
We refer to this parametrization as Model B. Using the parametrization
\eqref{eq:corass_parm}, on the other hand, we
need to fix values of the parameters $\Delta_{\rm m}$, $a_{\rm c}$ and $w_{\rm
m}$ in order to give the desired value of
$\Omega_{e}^{e}$. In this case the DE energy density is given by:
\begin{equation}
\rho_{\rm e}\left(a\right)=
\rho_{\rm
e}^{0}\exp\left(-3\int_{a_{0}}^{a}\frac{\left(1+w\left(a'\right)\right)da'}{a'}
\right)\,.
\end{equation}

\begin{figure}[tbp]
\centering
\includegraphics[scale=0.7]{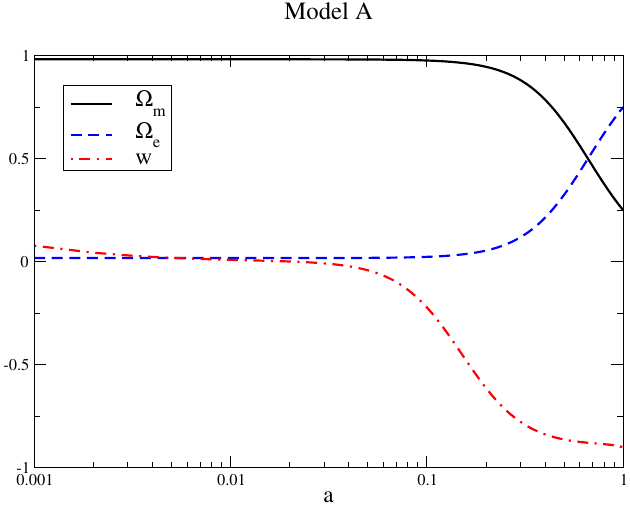}
\includegraphics[scale=0.7]{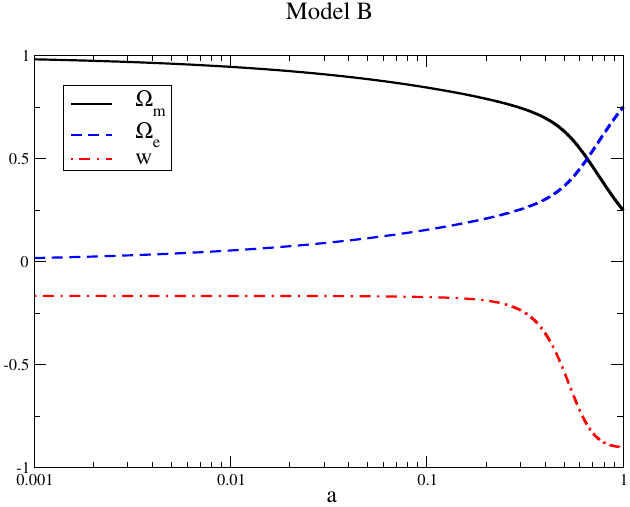}
\par
\caption{Evolution of matter and DE density
parameters, $\Omega_{\rm m}\left(a\right)$ (solid black line) and
$\Omega_{\rm de}\left(a\right)$ (blue dashed line), DE equation of state
$w\left(a\right)$ (red dot-dashed line). Left
(right) panel: Model A (B).}
\label{fig:background}
\end{figure}

For both models we choose the amount of DE at early times to be $\Omega_{\rm
e}^{\rm e}\simeq 0.018$, consistent with
the limits presented in \cite{Reichardt2012}, the amounts of matter and DE
now are $\Omega_{\rm m}^{0}=0.25$ and
$\Omega_{\rm e}^{0}=0.75$ and the DE equation of state now is $w_{0}=-0.9$. In
Model B we also set $w_{\rm m}=-0.1655$,
$a_{\rm c}=0.5$ and $\Delta_{\rm m}=0.09$. For these two models we show the
evolution of $\Omega_{\rm m}\left(a\right)$,
$\Omega_{\rm e}\left(a\right)$, $w\left(a\right)$ in Fig.~\ref{fig:background}.

As we can see in Fig.~\ref{fig:background}, in Model A, $\Omega_{\rm
e}\left(a\right)$ is basically constant at
high-$z$ and its equation of state varies slowly, whereas in Model
B the amount of DE has a non-negligible
variation during most of the cosmic time and has a rapid transition of its
equation of state for $a\simeq 1$. We will show that
these differences in the background evolution will imprint distinct features
both in matter and in DE fluctuations.
Assuming a Hubble constant of $H_0=72\, \mbox{km\,s}^{-1}\mbox{Mpc}^{-1}$ the
age of the universe in a $\Lambda$CDM model with $\Omega_{\rm m}^{0}=0.25$ is
$13.77$~Gy, whereas in Model A we have $13.44$~Gy and in Model B $13.01$~Gy.

\begin{figure}[tbp]
\centering
\includegraphics[scale=0.35]{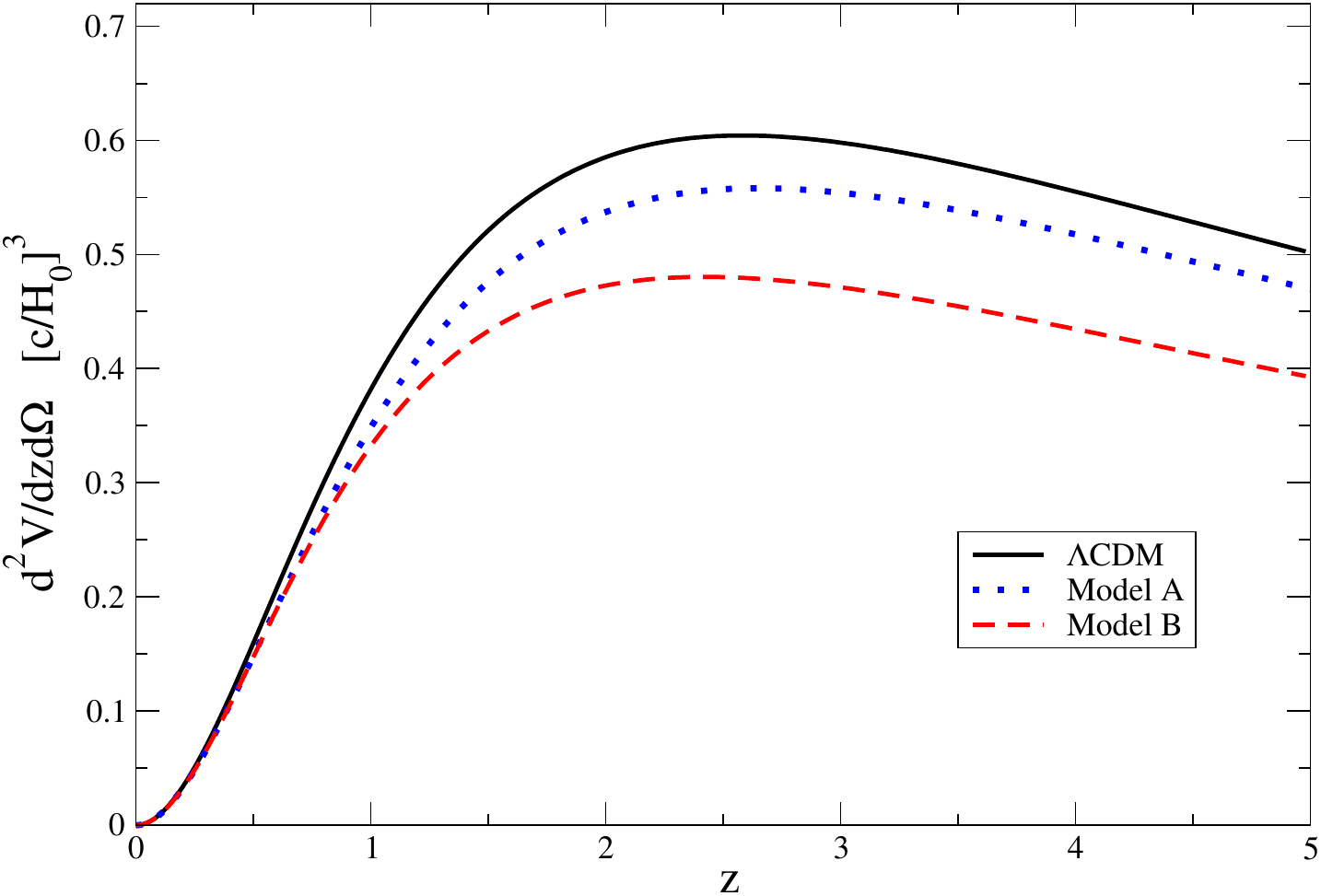}
\caption{Comoving volume for $\Lambda$CDM model (solid black line) and
models A
(blue dotted line) and B (red dashed line)
of EDE.}
\label{fig:volumes}
\end{figure}

In Fig.~\ref{fig:volumes} we also show the evolution of the comoving
volume, given by
\begin{equation}
\frac{d^{2}V}{dzd\Omega}=\frac{r^{2}\left(z\right)}{H\left(z\right)}\,,
\end{equation}
where $r\left(z\right)=\int_{0}^{z}H^{-1}\left(z'\right)dz'$. This quantity,
which depends only on the background evolution is important for the study of
cluster number counts, which in turn also depends on perturbative properties via
the mass function. Note that both models of EDE that we are considering present
a smaller volume then $\Lambda$CDM model and that important differences appear
only at high$-z$. For redshifts $z<0.5$ all three volumes are very similar,
which already suggests that cluster observations at low redshifts would poorly
differentiate between these DE models. We will return to this issue in Sect.
\ref{sect:mf}.

\section{Linear evolution}\label{sect:lin_evol}

In this section we study the linear relativistic evolution of a system with DE
and matter. In models of EDE the energy density of DE at high redshifts, e.g.,
at the redshift of decoupling $z_{\rm dec}$, is not negligible, hence we
carefully analyse approximate analytical solutions in order to establish
consistent initial conditions for the equations of motion. Moreover this study
will clarify some effects due to EDE and its perturbations.

In the Newtonian gauge, in Fourier space, in the absence of anisotropic stress,
Ref. \cite{Ma:1995ey}, the perturbed equations as a function of conformal time
can be written as: 
\begin{equation}
\dot{\delta}_{\rm m}+\theta_{\rm m}=3\dot{\phi}\;,
\label{comp_GR_mat_cont}
\end{equation}
\begin{equation}
\dot{\theta}_{\rm m}+\mathcal{H}\theta_{\rm m}=k^{2}\phi\;,
\label{comp_GR_mat_euler}
\end{equation}
\begin{equation}
\dot{\delta}_{\rm e}+3\mathcal{H}\left(\frac{\delta p_{\rm e}}{\delta\rho_{\rm
e}}-w\right)\delta_{\rm e}+
\left(1+w\right)\theta_{\rm e}=3\left(1+w\right)\dot{\phi}\;,
\label{comp_GR_de_cont}
\end{equation}
\begin{equation}
\dot{\theta}_{\rm e}+\mathcal{H}\left(1-3c_{\rm a}^{2}\right)\theta_{\rm e}=
\frac{\left(\delta p_{\rm e}/\delta\rho_{\rm e}\right)k^{2}\delta_{\rm
e}}{\left(1+w\right)}+k^{2}\phi\;,
\label{comp_GR_de_euler}
\end{equation}
\begin{equation}
k^{2}\phi+3\mathcal{H}\left(\dot{\phi}+\mathcal{H}\phi\right)=-
\frac{3\mathcal{H}^{2}}{2}\left(\Omega_{\rm m}\delta_{\rm m}+\Omega_{\rm
e}\delta_{\rm e}\right)\;,
\label{comp_PN_poisson}
\end{equation}
where
\begin{equation}
c_{\rm a}^{2}=\frac{\dot{p}_{\rm e}}{\dot{\rho}_{\rm
e}}=w-\frac{\dot{w}}{3\mathcal{H}\left(1+w\right)}
\label{eq:ad_sound}
\end{equation}
is the squared adiabatic sound speed and the pressure perturbation is given
by \cite{Bean2004}:
\begin{equation}
\delta p_{\rm e}=c_{{\rm eff}}^{2}\delta\rho_{\rm
e}+3\mathcal{H}\left(1+w\right)
\left(c_{{\rm eff}}^{2}-c_{\rm a}^{2}\right)\rho_{\rm e}\frac{\theta_{\rm
e}}{k^{ 2}}\,.
\label{eq:cov_press}
\end{equation}
Note that the equation of state can not cross the phantom barrier, otherwise
$c_{\rm a}^{2}$ will diverge, and so will DE perturbations. In the two models we
study there is no phantom crossing. For a treatment of this issue see
Ref.~\cite{Li2011}. We solve the system of equations
\eqref{comp_GR_mat_cont}-\eqref{comp_PN_poisson} numerically, however the
(ii)-Einstein equation, given by
\begin{equation}
\ddot{\phi}+3\mathcal{H}\dot{\phi}+\left(\frac{2\ddot{a}}{a}-\mathcal{H}^{2}
\right)\phi=
\frac{3\mathcal{H}^{2}}{2}\Omega_{\rm e}\left(\frac{\delta p_{\rm
e}}{\delta\rho_{\rm e}}\right)\delta_{\rm e}\,,
\label{phi_ddot_eq}
\end{equation}
will be useful to find analytical solutions.

It will also be useful to write a second order equation for the density
contrast. For the sake of simplicity we assume
that $c_{{\rm eff}}$ and $w$ are constants, then we have:
\begin{multline}
\ddot{\delta}+\left[\mathcal{H}\left(1-3w\right)-A\right]\dot{\delta}+
\left[3\Delta\left(\dot{\mathcal{H}}+\mathcal{H}^{2}\right)-3\mathcal{H}\Delta
A+c_{{\rm eff}}^{2}k^{2}\right]\delta
=\left(1+w\right)\left(S-k^{2}\phi\right)\,,
\label{delta_gen_ddot}
\end{multline}
where $\Delta=c_{{\rm eff}}^{2}-w$,
\begin{equation}
A=\frac{18\mathcal{H}\dot{\mathcal{H}}\Delta}{9\mathcal{H}^{2}\Delta+k^{2}}\,
\end{equation}
and
\begin{equation}
S=3\ddot{\phi}+3\mathcal{H}\left(1-3c_{{\rm
eff}}^{2}\right)\dot{\phi}-3A\dot{\phi}-9\mathcal{H}^{2}\Delta\phi\,.
\end{equation}
Eq.~\eqref{delta_gen_ddot} is valid for any perfect fluid characterised by
constant $w$ and $c_{{\rm eff}}^{2}$,
so the corresponding equation for matter is given by assuming $w=c_{{\rm
eff}}^{2}=0$.

The function $A$ is related to the presence of the pressure perturbations that
are proportional to $\theta$. It is
useful to note that on small scales, $k\gg\mathcal{H}$, $A\simeq 0$ and
$\delta p_{\rm e}\simeq c_{{\rm eff}}^{2}\delta\rho_{\rm e}$, as already
observed in \cite{Ballesteros2010b}. On large
scales, $k\ll\mathcal{H}$, we have $A\simeq2\dot{\mathcal{H}}/\mathcal{H}$ and
the pressure perturbation strongly
deviates from the value prescribed by $c_{{\rm eff}}$.

There are two important scales that determine the qualitatively behaviour of DE
perturbations: the particle horizon
\begin{equation}
\lambda_{H}\left(a\right)=\int_{a_{i}}^{a}\frac{da^{\prime}}{a^{\prime}\mathcal{
H}}\,,
\end{equation}
(horizon for short), which is also important for matter perturbations, and the
sound horizon
\begin{equation}
\lambda_{s}\left(a\right)=\int_{a_{\rm i}}^{a}\frac{c_{{\rm
eff}}\left(a\right)da^{\prime}}{a^{\prime}\mathcal{H}}\,.
\end{equation}
Perturbations with wavelength smaller than $\lambda_{s}$ oscillate with
decreasing amplitude and eventually reach a
minimum value proportional to the gravitational potential. Perturbations with
wavelength larger than $\lambda_{s}$
effectively behave as pressureless, growing at the same pace as matter
perturbations. Finally perturbations in both
matter and DE with wavelength larger than $\lambda_{H}$ will follow the time
dependence of the gravitational
potential, being constant whenever DE effect in the background is negligible.

We are interested in small scales were nonlinear structures, such as galaxy
clusters, form. Typically we can assume
that the order of magnitude of such scales is $10\,\mbox{h}^{-1}\,\mbox{Mpc}$,
or
$k_{\rm nl}\simeq 0.63\,\mbox{h}\,\mbox{Mpc}^{-1}$. Shortly after decoupling,
e.g., at $z=1000$, the horizon wave
number is $k_{H}\simeq 0.017\,\mbox{h}\,\mbox{Mpc}^{-1}$ ($\Lambda$CDM with
$\Omega_{\rm m}^{0}=0.25$ background is
assumed), hence both DE and DM perturbations that go nonlinear are well inside
the horizon on the onset of matter
dominated period. For this reason we will focus our analysis on scales much
smaller than $\lambda_{H}$, consequently the
only scale important for DE perturbations is $\lambda_{s}$. For a study of
scales larger than the horizon
see Ref.~\cite{Ballesteros2010b}.

It is interesting to determine the value of $c_{{\rm eff}}$, assumed constant,
such that $k_{s}=k_{\rm nl}$ today:
\begin{equation}
c_{\rm nl}=9.4\times10^{-4}\,.
\end{equation}
In models with $c_{{\rm eff}}\ll c_{\rm nl}$ the sound horizon is always smaller
than the nonlinear scale, then DE
perturbations will behave as pressureless and we can effectively assume $c_{{\rm
eff}}=0$. On the other hand, in models
with $c_{{\rm eff}}\gg c_{\rm nl}$, DE perturbations are suppressed by its
pressure support on nonlinear scales. In this
work we treat the two limiting cases: one with negligible sound speed, $c_{{\rm
eff}}=0$, and one with a non-negligible sound speed, $c_{{\rm eff}}=1$. For
studies of time dependent $c_{{\rm eff}}$ in linear theory see
Ref.~\cite{Ansari2011} and for arbitrary values of $c_{{\rm eff}}$  during
nonlinear evolution of DM see Ref.~\cite{Basse2011,Basse2012}.

\subsection{Non-negligible \textmd{\normalsize $c_{{\rm eff}}$}}

On scales well inside the horizon, for constant $c_{{\rm eff}}$ and $w$, from
Eq.~\eqref{delta_gen_ddot}, DE perturbations obey the following equation:
\begin{equation}
\delta_{e}^{\prime\prime}+\alpha\frac{\delta_{e}^{\prime}}{a}+
\left(\beta+\frac{c_{{\rm
eff}}^{2}k^{2}}{\mathcal{H}^{2}}\right)\frac{\delta_{e}}{a^{2}}=
-\frac{\left(1+w\right)}{\mathcal{H}^{2}}\frac{k^{2}\phi}{a^{2}}\,,
\label{delta_gen_a}
\end{equation}
where $^{\prime}=d/da$, $w_{\rm t}=\Omega_{\rm e}w$,
\begin{equation}
\alpha=\frac{3}{2}\left(1-w_{\rm t}\right)-3w\,,\,\,\,\,\,\mbox{and}\,\,\,
\beta=\frac{3\Delta}{2}\left(1-3w_{\rm t}\right)\,.
\end{equation}
Eq.~\eqref{delta_gen_a} has the particular approximate solution:
\begin{equation}
\delta_{\rm e}\simeq-\frac{\left(1+w\right)\phi}{c_{{\rm eff}}^{2}}\,,
\label{delta_part}
\end{equation}
which becomes more accurate the bigger $c_{{\rm eff}}k/\mathcal{H}$ is. Note
that since $c_{{\rm eff}}k/\mathcal{H}$
grows in time, as long as the background parameters do not change too fast in
time, $\delta_{\rm e}$ will approach
solution \eqref{delta_part} on all scales. Moreover, the homogeneous solutions
of Eq.~\eqref{delta_gen_a} oscillate with
decreasing amplitude \cite{Abramo:2008ip,Ballesteros2010b}, hence eventually
solution \eqref{delta_part} will be reached
and can be used to set the initial value of $\delta_{\rm e}$  as a
function of $\phi$.

Before we determine the initial value of $\theta_{\rm e}$, we proceed to find
solutions for $\phi$. Since we are
interested on small scales we can assume that $\delta p_{\rm e}\simeq c_{{\rm
eff}}^{2}\delta\rho_{\rm e}$. We verified
that, for the IC we will determine and $c_{{\rm eff}}=1$, the error in this
expression compared to
Eq.~\eqref{eq:cov_press} for a scale $k=0.1\mbox{h}\mbox{Mpc}^{-1}$ is initially
of $1-3\%$ and below $1\%$ for $z<100$
and that the error decreases for smaller scales. Then, substituting the
expression \eqref{delta_part} in
Eq.~\eqref{phi_ddot_eq} and changing the independent variable to the scale
factor $a$, the evolution of the
gravitational potential is given by the equation:
\begin{equation}
\phi^{\prime\prime}+\left(\frac{7}{2}-\frac{3}{2}w_{\rm
t}\right)\frac{\phi^{\prime}}{a}+
\left(\frac{3\left(1+w\right)}{2}\Omega_{\rm e}-3w_{\rm
t}\right)\frac{\phi}{a^{2}}=0\,.
\label{eq:phi_ddot_solve}
\end{equation}
This equation is valid as long as $\Omega_{\rm e}$ and $w$ are nearly constant,
which is true for Model A roughly for
$10<z<1000$ and for Model B roughly for $100<z<1000$. Solving this equation in
this interval enables us to analytically evaluate the time variation of $\phi$.
The solutions of Eq.~\eqref{eq:phi_ddot_solve} are power laws
$\phi\propto a^{n}$, where $n$ is solution of the algebraic equation
\begin{equation}
n^{2}+\left(\frac{5}{2}-\frac{3}{2}w_{\rm t}\right)n+\frac{3\left(1+w\right)
\Omega_{\rm e}}{2}-3w_{\rm t}=0\,.\label{power_phi}
\end{equation}
For non-EDE models, when DE is very subdominant, we can set $\Omega_{\rm e}=0$
and the solutions are $n=0$ and
$n=-2.5$, which indicates that, neglecting the decaying mode, the gravitational
potential is constant during matter
dominated era, a well known result for the Einstein-de-Sitter universe. However,
in EDE models $\phi$ is not constant
any more, its early time variation depends on both $\Omega_{\rm e}$ and $w$, but
note it is independent of $c_{{\rm
eff}}$.

The largest values of $n$, which give the solutions that decay more slowly, are
\begin{equation}
n_{\rm mod~A}=-0.0100\,\mbox{ and }\,n_{\rm mod~B}=-0.0127.
\label{eq:dphi_A}
\end{equation}
Since in Model A both $\Omega_{\rm e}$ and $w$ are approximately constant before
DE domination, the error in this
expression against the full numerical solution is below $1\%$ in the range
$5<z<1000$. Given that in Model B the time
variation of $\Omega_{\rm e}$ and $w$ is larger, its analytical solution has an
error below $1\%$ only in the range
$150<z<1000$.

Once we know how $\phi$ initially varies in time we can find $\dot{\delta}_{\rm
e}$ from Eq.~\eqref{delta_part} and use
Eq.~\eqref{comp_GR_de_cont} to determine the initial value of $\theta_{\rm e}$:
\begin{equation}
\theta_{\rm e}=3\dot{\phi}-\frac{\dot{\phi}}{c_{{\rm eff}}^{2}}-
\frac{3\mathcal{H}\left(c_{{\rm eff}}^{2}-w\right)\phi}{c_{{\rm eff}}^{2}}\,.
\label{theta_delta_e_IC}
\end{equation}
Hence Eq.~\eqref{theta_delta_e_IC} can be used to determine the initial value of
$\theta_{\rm e}$ as a function of the
initial value of $\phi$ and its initial time derivative.

Finally we look for analytical solutions for matter perturbations in order to
determine IC for $\delta_{\rm m}$ and
$\theta_{\rm m}$. Setting $w=c_{{\rm eff}}=0$ in Eq.~\eqref{delta_gen_a} we
have:
\begin{equation}
\delta_{\rm m}^{\prime\prime}+\left(\frac{3}{2}-\frac{3w_{\rm
t}}{2}\right)\frac{\delta_{\rm m}^{\prime}}{a}=
-\frac{k^{2}\phi}{\mathcal{H}^{2}a^{2}}
\label{delta_m_ddot_a}
\end{equation}
It is useful to note that, although in EDE models $\Omega_{\rm e}$ is
non-negligible at high-$z$, on small scales, since
$\delta_{\rm e}\sim\phi$, we can assume that $\Omega_{\rm m}\delta_{\rm
m}\gg\Omega_{\rm e}\delta_{\rm e}$ in
Eq.~\eqref{comp_PN_poisson}, then we have:
\begin{equation}
\delta_{\rm m}\simeq\frac{-2k^{2}\phi}{3\mathcal{H}^{2}\Omega_{\rm m}}\,.
\label{phi_delta_m_IC}
\end{equation}
Using Eq.~\eqref{phi_delta_m_IC} we can find the initial value of $\delta_{\rm
m}$ as a function of the initial value of a
given $\phi$. Inserting Eq.~\eqref{phi_delta_m_IC} in Eq.~\eqref{delta_m_ddot_a}
we have the equation for $\delta_{\rm
m}$ alone:
\begin{equation}
\delta_{\rm m}^{\prime\prime}+\left(\frac{3}{2}-\frac{3w_{\rm
t}}{2}\right)\frac{\delta_{\rm m}^{\prime}}{a}-
\frac{3\Omega_{\rm m}}{2}\frac{\delta_{\rm m}}{a^{2}}=0\,,
\label{delta_m_eq_c1}
\end{equation}
which is the usual equation for the evolution of matter perturbations well
inside the horizon. The solutions are power
laws $\delta_{\rm m}\propto a^{p}$, where $p$ is determined by:
\begin{equation}
p^{2}+\left(\frac{1}{2}-\frac{3w_{\rm t}}{2}\right)p-\frac{3\Omega_{\rm
m}}{2}=0\,.
\label{p_eq_c}
\end{equation}
Note that, under the approximations taken, when we compare Eqs.~\eqref{p_eq_c}
and \eqref{power_phi}, we find that
$n=p-1$. Then the corresponding values for $p$ are:
\begin{equation}
p_{\rm mod~A}=0.9900\,\,\,\mbox{and}\,\,\,\, p_{\rm mod~B}=0.9873\,.
\label{p_c1}
\end{equation}

In order to find the initial value of $\theta_{\rm m}$ we can use
Eq.~\eqref{comp_GR_mat_cont}. The initial value of
$\delta_{\rm m}$ and its initial dependence with $a$, is given by finding $p$ in
Eq.~\eqref{p_eq_c}. Then on small scales we have:
\begin{equation}
\theta_{\rm m}=-\dot{\delta}_{\rm m}\label{theta_m_IC}\,.
\end{equation}

\begin{figure}[tbp]
\centering
\includegraphics[scale=0.65]{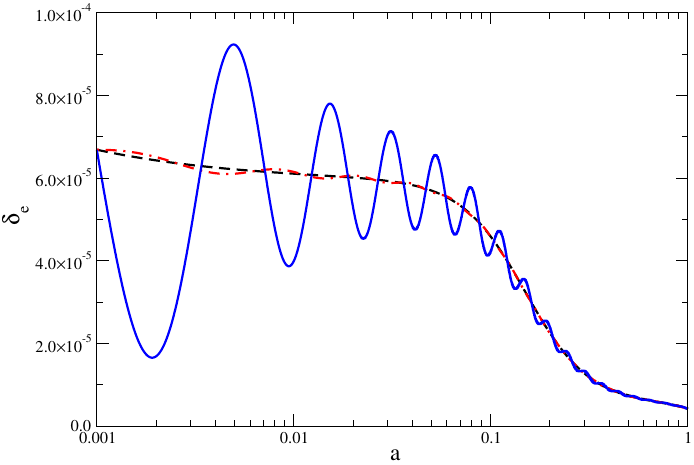}
\includegraphics[scale=0.65]{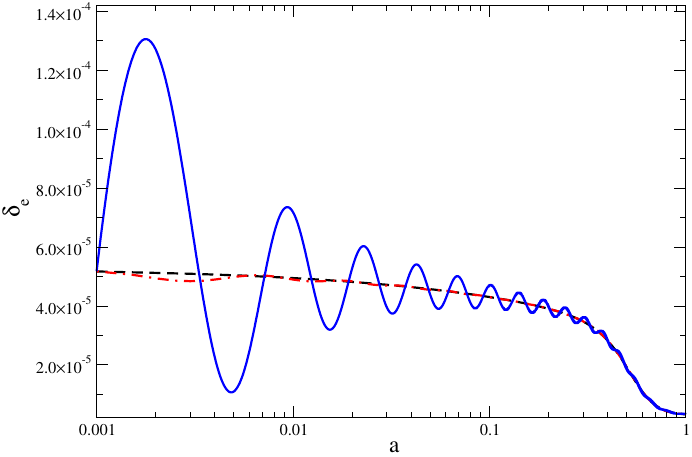}
\caption{Linear evolution of $\delta_{\rm e}$ with $c_{{\rm eff}}^{2}=0.1$ for a
mode $k=0.1\mbox{h}\mbox{Mpc}^{-1}$:
analytical solution, Eq.~\eqref{delta_part} (black dashed line), numerical
solution with IC from Eqs.~\eqref{delta_part}
and \eqref{theta_delta_e_IC} (red dot-dashed line) and the numerical solution
with $\theta_{\rm e}$ ten times bigger
(blue line). The change in the analytical solution due to the different IC is
almost imperceptible in this scale so we
only show the first case. }
\label{fig:deltas_DR}
\end{figure}

In Fig.~\ref{fig:deltas_DR} we show the evolution of $\delta_{\rm e}$ according
to the numerical solution of
equations~\eqref{comp_GR_mat_cont}$-$\eqref{comp_PN_poisson}, and the analytical
solution of Eq.~\eqref{delta_part}, with $\phi$ given by the numerical
solution. The only value we need to choose
is $\phi_{i}=\phi\left(a_{i}\right)$, taking $a_{i}=0.001$ and
$\phi_{i}=-6.2\times10^{-6}$ we	get $\delta_{m}\left(1\right)\simeq 0.1$. In
order to check the consistency of the
analytical solutions that we have used to determine the IC of DE perturbations,
given by Eq.~\eqref{delta_part} and Eq.~\eqref{theta_delta_e_IC}, (red
dotted-dashed line) we also show the evolution
of $\delta_{\rm e}$ using $\theta_{\rm e}$ ten times greater than as given by
Eq.~\eqref{theta_delta_e_IC} (blue line). As we see, the solution of
Eq.~\eqref{delta_part} is in good agreement with the numerical solution. In the
case of bigger $\theta_{\rm e}$, $\delta_{\rm e}$ oscillates with an initial
greater amplitude, which decreases with time and eventually approaches the value
given by Eq.~\eqref{delta_part}. For sake of clarity in these plots, we assumed
$c_{{\rm eff}}^{2}=0.1$ in order to have a smaller frequency of oscillation. It
is important to note that for this scale and for this value of $c_{\rm eff}$, at
low $z$, DE perturbations are $4$ orders of magnitude smaller then $\delta _m$.
For smaller scales and larger $c_{\rm eff}$, $\delta _{\rm e}$ will be even
smaller
than $\delta _{\rm m}$.

\subsection{Negligible $c_{{\rm eff}}$ }

As we will see, in the case of negligible $c_{{\rm eff}}$, DE perturbations can
have the same order of magnitude of
matter perturbations. Although this fact impedes us to assume that $\Omega_{\rm
m}\delta_{\rm m}\gg\Omega_{\rm
e}\delta_{\rm e}$ for EDE models, fact that simplified the equations for the
case of non-negligible
$c_{{\rm eff}}$, when $c_{{\rm eff}}$ is negligible we will have another sort of
simplification. Note that substituting
the expression for the pressure perturbation, Eq.~\eqref{eq:cov_press}, in the
equation for $\theta_{\rm e}$,
Eq.~\eqref{comp_GR_de_euler}, and assuming $c_{{\rm eff}}=0$ we obtain

\begin{equation}
\dot{\theta}_{\rm e}+\mathcal{H}\theta_{\rm e}=k^{2}\phi\,,
\end{equation}
which is just the same equation for $\theta_{\rm m}$,
Eq.~\eqref{comp_GR_mat_euler}. Hence both matter and DE
perturbations will feel the same force and flow in the same way. Then, using
Eqs.~\eqref{comp_GR_mat_cont} and
\eqref{comp_GR_de_cont}, we can determine an equation that directly relates
$\delta_{\rm m}$ and $\delta_{\rm e}$:
\begin{equation}
\dot{\delta}_{\rm e}-3\mathcal{H}w\delta_{\rm
e}=\left(1+w\right)\dot{\delta}_{\rm m}\,,
\end{equation}
where, since, on small scales, both $\delta_{\rm m}$ and $\delta_{\rm e}$ will
be of order
$\mathcal{H}^{-2}k^{2}\phi$, we neglected the
$\dot{\phi}$ term. Assuming that $\delta_{\rm m}\propto a^{p}$ we find the
following relation between DE and
matter perturbations:
\begin{equation}
\delta_{\rm e}=\frac{\left(1+w\right)p}{p-3w}\delta_{\rm m}\;.
\label{de_dm_sol_gen}
\end{equation}
In models with negligible amount of DE during matter era we have $p=1$, then
\begin{equation}
\delta_{\rm e}=\frac{\left(1+w\right)}{\left(1-3w\right)}\delta_{\rm m}\,,
\label{de_dm_old_sol}
\end{equation}
which is in accordance with literature
\cite{Abramo:2008ip,Sapone2009,Creminelli2010}. However, as we will show in
Eq.~\eqref{p_c0}, $p\simeq 1$ in EDE models with negligible $c_{{\rm eff}}$,
then we can actually use Eq.~\eqref{de_dm_old_sol} to set the initial value of
$\delta_{\rm e}$ as a function of $\delta_{\rm m}$. It is also interesting to
observe that Eq. \eqref{de_dm_old_sol} indicates that DE perturbations diverge
as $w\rightarrow 1/3$, which can be achieved by a tracking scalar field during
radiation dominated era. However this solution is valid only for $c_{{\rm
eff}}=0$, hence, as consistency condition for DE perturbations, we should expect
that whenever $w\rightarrow 1/3$ we must have $c_{{\rm eff}}>0$. In
the models we are studying, during the matter dominated era and DE domination,
we never have $w=1/3$.

At this point we derived a relation between DE and matter perturbations but not
their evolution with time. Setting
$w=c_{{\rm eff}}=0$ in Eq.~\eqref{delta_gen_a} we have the equation for the
evolution of $\delta_{\rm m}$:
\begin{equation}
\delta_{\rm m}^{\prime\prime}+\left(\frac{3}{2}-\frac{3w_{\rm t}}{2}\right)
\frac{\delta_{\rm m}^{\prime}}{a}=
\frac{3}{2a^{2}}\left(\Omega_{\rm m}\delta_{\rm m}+\Omega_{\rm e}\delta_{\rm
e}\right)\,.
\label{delta_m_eq1}
\end{equation}
Now we can use Eq.~\eqref{de_dm_sol_gen} to express $\delta_{\rm e}$ as a
function of $\delta_{\rm m}$ and rewrite
Eq.~\eqref{delta_m_eq1} as

\begin{equation}
\delta_{\rm m}^{\prime\prime}+\left(\frac{3}{2}-\frac{3w_{\rm t}}{2}\right)
\frac{\delta_{\rm m}^{\prime}}{a}-
\frac{3}{2}\left(\Omega_{\rm m}+\frac{\left(1+w\right)p}{p-3w}\Omega_{\rm
e}\right)
\frac{\delta_{\rm m}}{a^{2}}=0\,.\label{delta_m_eq2}
\end{equation}
While $\Omega_{\rm m}$, $\Omega_{\rm e}$ and $w$ are nearly constants we can
find the solution
$\delta_{\rm m}\propto a^{p}$, where $p$ is solution of
\begin{equation}
p^{2}+p\left(\frac{1}{2}-\frac{3w_{\rm
t}}{2}\right)-\frac{3}{2}\left(\Omega_{\rm m}
+\frac{\left(1+w\right)p}{p-3w}\Omega_{\rm e}\right)=0\,.\label{p_eq_c_zero}
\end{equation}
For the two models we are studying we have the following values at $a_i$:
\begin{equation}
p_{\rm mod~A}=1.0052\,\,\,\mbox{and}\,\,\,\, p_{\rm mod~B}=0.9934\,.\label{p_c0}
\end{equation}
As we can see, the deviation from $p=1$ is very small and we could actually use
expression \eqref{de_dm_old_sol} in
order to determine the initial value of $\delta_{\rm e}$ as a function of
$\delta_{\rm m}$. Moreover expression
\eqref{de_dm_old_sol} gives a better order of magnitude estimation of
$\delta_{\rm e}$ for low$-z$ than
Eq.~\eqref{de_dm_sol_gen}, that depends on the determination of $p$ from
Eq.~\eqref{p_eq_c_zero}, which is not
accurate for low$-z$, when both $\Omega_{\rm e}$ and $w$ can not be assumed
constant. Hence in what follows we
will use the expression \eqref{de_dm_old_sol} to compute the initial value of
the DE perturbation.

To determine the initial value of $\delta_{\rm m}$ as a function of $\phi$ we
use Eq.~\eqref{de_dm_sol_gen} in
Eq.~\eqref{comp_PN_poisson}, then, for small scales we have:
\begin{equation}
\delta_{\rm m}=-\frac{2k^{2}\phi}{3\mathcal{H}^{2}\left(\Omega_{\rm
m}+\Omega_{\rm e}\frac{1+w}{1-3w}\right)}\,.
\label{dm_IC_c_zero}
\end{equation}
The initial value of $\theta_{\rm m}$ can be determined in the same way as we
did for the case of non-negligible
$c_{{\rm eff}}$, using Eq.~\eqref{theta_m_IC}.

Now we proceed to find the initial time variation for $\phi$. For the case
$c_{{\rm eff}}=0$ DE perturbations do not
contribute to the variation of $\phi$, i.e., the RHS of Eq.~\eqref{phi_ddot_eq}
is zero and then the equation that
determines the power of $n$ of $\phi$ is the following:
\begin{equation}
n^{2}+n\left(\frac{5}{2}-\frac{3}{2}w_{\rm t}\right)-3w_{\rm
t}=0\,.\label{n_phi_c_zero}
\end{equation}
Here, in contrast to the case of non-negligible $c_{{\rm eff}}$, the relation
$n=p-1$ is not valid. Note also that $n$
only depends on the product $\Omega_{\rm e}w$, differently from case of
non-negligible $c_{{\rm eff}}$, where it
also depends directly on $\Omega_{\rm e}$. This indicates that EDE models with
negligible sound speed give a very low
contribution to the time variation of $\phi$, due to the fact that, although
$\Omega_{\rm e}$ is not negligible during
the matter era, we have $\Omega_{\rm e}w\simeq 0$. The highest values of $n$ for
models A and B are:
\begin{equation}
n_{\rm mod~A}=0.0017\,\,\,\mbox{and}\,\,\: n_{\rm
mod~B}=-0.0036\,.\label{eq:dphi_A_c0}
\end{equation}
Note that the absolute value of $n$ is one order of magnitude smaller than the
solutions for the case of non-negligible
$c_{{\rm eff}}$, Eq.~\eqref{eq:dphi_A}. Since the background evolution is the
same for both cases, this clearly shows
that the most important effect for the evolution of the gravitational potential
in EDE models is pressure perturbation.

\begin{figure}[tbp]
\centering{}
\includegraphics[scale=0.65]{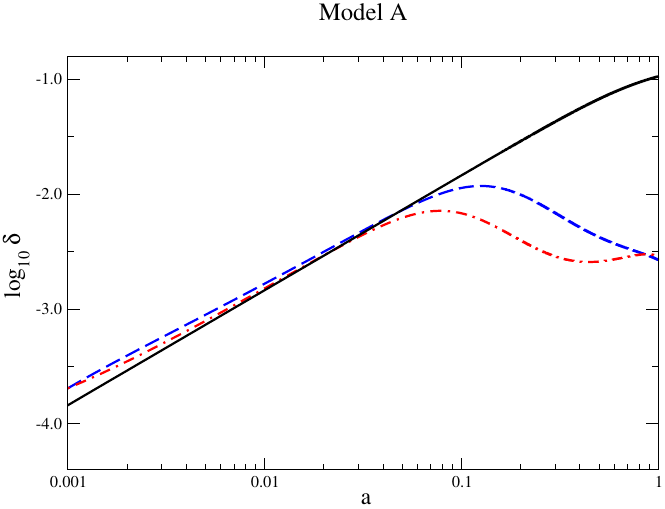}
\includegraphics[scale=0.65]{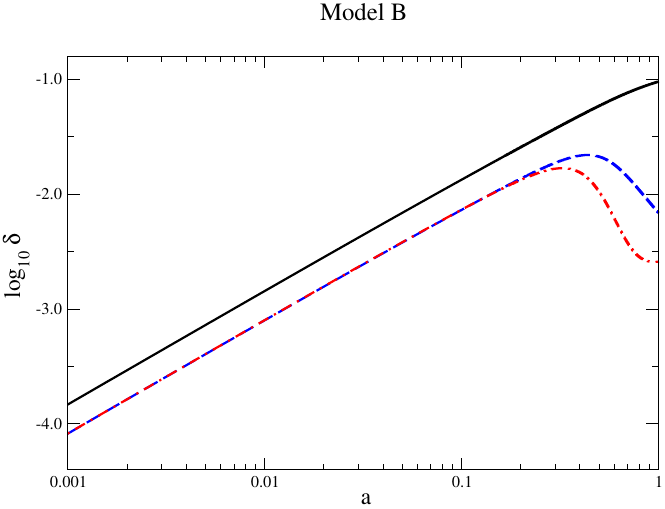}
\caption{Linear evolution of perturbations with $c_{{\rm eff}}=0$ and
$k=0.1\mbox{h}\mbox{Mpc}^{-1}$: numerical
solution of $\delta_{\rm m}$ (black solid line), numerical solution of
$\delta_{\rm e}$ (blue dashed line) and
$\delta_{\rm e}$ according to the relation given by the analytical solution,
Eq.~\eqref{de_dm_old_sol},
(red dot-dashed line).}
\label{fig:deltas_c0}
\end{figure}

In Fig.~\ref{fig:deltas_c0} we show the evolution of matter and DE perturbations
as well as the evolution of
$\delta_{\rm e}$ according to Eq.~\eqref{de_dm_old_sol}. We show this relation,
instead of the more general one, in
Eq.~\eqref{de_dm_sol_gen}, because initially the deviation from $p=1$ is very
small (bellow $1\%$) and for late times,
when $\Omega_{\rm m}$, $\Omega_{\rm e}$ and $w$ are not constant, the
computation of $p$ via Eq.~\eqref{p_eq_c_zero} is
less accurate than just assuming $p=1$. As we can see, Eq.~\eqref{de_dm_old_sol}
is a very good approximation for
$\delta_{\rm e}$ during the matter dominated era, however, once the transition
to the DE dominated era starts, it
becomes much less accurate. Anyhow Eq.~\eqref{de_dm_old_sol} can be used as an
estimate of the order of magnitude of
$\delta_{\rm e}$ for late times. It is also interesting to note that in Model A
the perturbations in DE are larger
than matter perturbations until $w$ starts to decrease, $a\simeq 0.1$. For Model
B, since $1+w$ is smaller than in Model
A, $\delta_{\rm e}$ is smaller during most of the evolution. However, due to the
latter transition of $w$ in Model B,
$\delta_{\rm e}$ starts to decrease at a later time and its final value is
larger then in Model A.

\subsection{Matter growth and ISW effect}

Once we know approximate analytical solutions to set the IC of the system of
equations, we proceed to study the effects of EDE models on the linear evolution
of matter perturbations and gravitational potential.

\begin{figure}[tbp]
\centering
\includegraphics[scale=0.6]{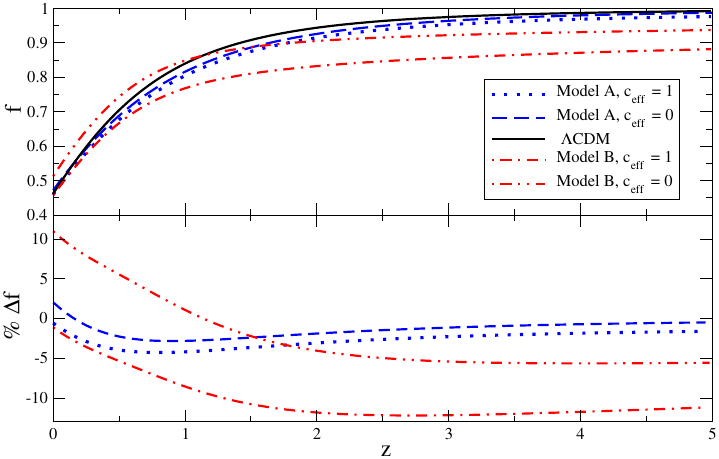}
\includegraphics[scale=0.6]{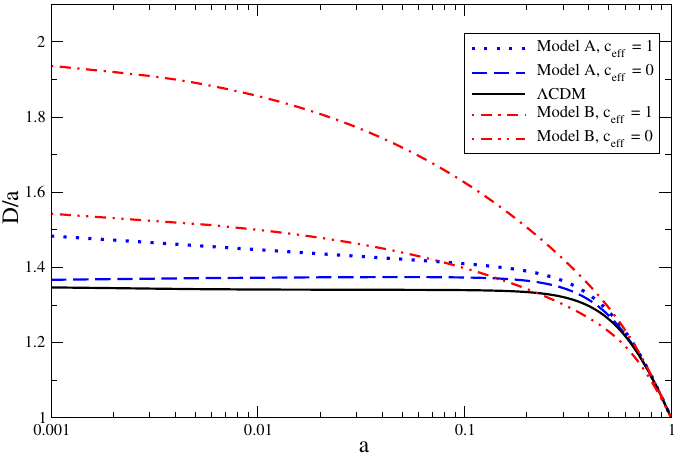}
\caption{Evolution of $f$ (left top panel), its percentage difference against
$\Lambda$CDM model (left bottom panel) and the normalised growth factor $D/a$
(right panel), for a mode $k=0.1\mbox{h}\mbox{Mpc}^{-1}$. Line styles and
colours for the different models are indicated in the legends.
}
\label{fig:grow_DM_c}
\end{figure}

In Fig.~\ref{fig:grow_DM_c} we show the evolution of the normalised growth
function $D/a$, where $D=\delta_{\rm
m}\left(a\right)/\delta_{\rm m}\left(1\right)$, and the logarithmic derivative
of the growth function $f=d\mbox{ln}D/d\mbox{ln}a$ and its percentual
difference against $\Lambda$CDM,
\begin{equation}
\Delta f = 100\times \frac{f_{\rm mod}-f_{\Lambda CDM}}{f_{\Lambda CDM}}\, .
\end{equation}
In the $\Lambda$CDM model, both the normalised
growth and $f$ tend to a constant value
for low $a$ (high $z$) because $D\rightarrow a$ whenever DE is very
subdominant. In this limit $f\rightarrow1$ and
$D/a\rightarrow1.34$. 

In EDE models $D/a$ and $f$ vary at high $z$, but in the
case of $c_{{\rm eff}}=0$, DE perturbations
tend to compensate the change in background evolution due to EDE and their
values get closer to the $\Lambda$CDM values. In the case of $c_{{\rm eff}}=1$,
DE perturbations are much smaller and practically do not
compensate the change in the background evolution, thus the deviation from
$\Lambda$CDM is more prominent.

This behavior can be clearly understood in terms of Eqs.
\eqref{delta_m_eq_c1} and \eqref{delta_m_eq1}. For a given model, when $c_{{\rm
eff}}=0$, DE perturbations enhance DM clustering via the last term in the LHS of
Eq. \eqref{delta_m_eq1}, which is absent in Eq. \eqref{delta_m_eq_c1}. Since EDE
models have a less decelerated background expansion than $\Lambda$CDM
model during the matter dominated era, which in turn makes DM clustering less
efficient, the contribution of DE perturbations in the case of $c_{{\rm eff}}=0$
tend to compensate this change, making DM growth in these models more similar to
the $\Lambda$CDM one at high $z$.

At low $z$ the situation is reversed, EDE models show a less accelerated
expansion, which induces a faster DM growth than in $\Lambda$CDM. Again DE
perturbations enhance this growth, but now making it more different from the
$\Lambda$CDM one. Therefore, as we can see in Fig.~\ref{fig:grow_DM_c},  $\Delta
f$ becomes larger at low $z$, particularly for EDE models with $c_{{\rm eff}}=0$.
Note that this effect is more important in Model B, which presents larger DE
perturbations at low $z$. However, the integrated impact of DE perturbations,
which can be observed via $D/a$ at $a=0.001$, always makes DM growth more
similar to $\Lambda$CDM. As we will see in Sect. \ref{sect:mf} the integrated
impact of EDE models in DM growth will cause large differences in
the abundance of galaxy clusters via the normalisation of the matter power
spectrum, $\sigma_8$, but the specific influence of DE pertubations is to make
cluster abundace more similar to the predictions of $\Lambda$CDM model.     

Besides this general behaviour of EDE models, we
can see that Model B produces the most distinct evolution and variation between
the two values of $c_{{\rm eff}}$. This is mainly caused by the rapid transition
of its equation of state, which produces a very different background evolution
and also allows DE perturbations to grow for a longer period when 
$c_{{\rm eff}}=0$. Although the background evolution of Model B may be
unrealistic, for instance, providing a rather low age of $13.01\,$Gy, it is very
valuable to clearly identify the changes that large DE perturbations may cause.
Since current data on $f$ have a precision around $10\%$ \cite{Blake:2011rj},
even in Model B, large DE perturbations would be weakly distinguished from the
case of negligible DE fluctuations. However future surveys like Euclid, which
may achieve about $1\%$ precision \cite{Amendola:2012ys}, could make a much more
significant distinction.

\begin{figure}[tbp]
\centering
\includegraphics[scale=0.75]{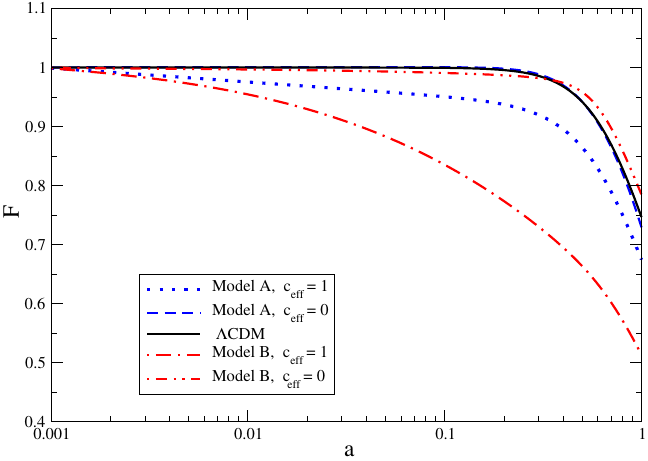}
\caption{Evolution of the function $F$,  Eq.~\eqref{phi_F}, as a function of
the scale factor. The curve of Model A with $c_{{\rm eff}}=0$ is almost
indistinguishable from $\Lambda$CDM model.
}
\label{fig:ISW}
\end{figure}

The Sachs-Wolfe effect \cite{Sachs:1967er} causes the gravitational blueshift
(redshift) of CMB photons when they fall in (escape from) potential wells near
the time of last scattering. When the gravitational potential varies in time,
the temperature variation of CMB photons is associated with the Integrated
Sachs-Wolfe effect (ISW). Assuming zero opacity the temperature fluctuation is
proportional to:
\begin{equation}
\frac{\Delta T}{T}\propto\int_{\eta_{0}}^{\eta_{\rm dec}}\dot{\phi}d\eta=
\phi\left(\eta_{\rm dec}\right)-\phi\left(\eta_{0}\right)\,,
\end{equation}
where $\eta_{\rm dec}$ and $\eta_{0}$ are the conformal time at decoupling
$\left(z_{\rm dec}\simeq 1100\right)$ and
now, respectively. The ISW effect can be separated into two components: the
early ISW, which is usually attributed to
the influence of residual radiation energy after the redshift of last scattering
and the late ISW, associated with the
accelerated expansion at low redshift. Since we neglect radiation, we will
only deal with ISW originated by DE. While
$\Omega_{\rm m}\simeq1$, the gravitational potential is nearly constant on all
scales and the corresponding ISW during
matter dominated era is negligible. However, for EDE models we have $\Omega_{\rm
m}\simeq 0.99$ at high$-z$, so the background evolution alone should generate a
distinct time evolution for $\phi$. Moreover, pressure perturbations in EDE
directly source the time variation of the gravitational potential,
Eq.~\eqref{phi_ddot_eq}, and it turns out that this
contribution is more important. In order to visualise how EDE and its
perturbations generate the ISW effect, in Fig.~\ref{fig:ISW} we show the
function
\begin{equation}
F=\frac{\phi\left(a\right)}{\phi\left(a_{dec}\right)}\,,
\label{phi_F}
\end{equation}
so we have $\frac{\Delta T}{T}\propto\phi\left(a_{\rm
dec}\right)\left(1-F\left(1\right)\right)$. One can clearly see
that the time variation of $\phi$ is greater in models with $c_{{\rm eff}}=1$
than $c_{{\rm eff}}=0$. Since the background evolution of Model A is similar to
$\Lambda$CDM, in the case of $c_{{\rm eff}}=0$ the ISW effect is also very
similar to the $\Lambda$CDM model. Again, the general effect of models with null
$c_{{\rm eff}}$ is to compensate the changes of EDE in the background
evolution. Although in principle ISW has limited power to constrain DE models
due to its large cosmic variance, the observation of this effect is improving
and a detection of $4.4\sigma$ level has been reported
\cite{Giannantonio:2012aa}.

\section{Nonlinear evolution}\label{sect:nl_evol}

Now we proceed to study the nonlinear evolution of DE and matter fluctuations,
which can provide important information about the formation of DM halos. One
difficulty that arises in this analysis is how to evolve DE fluctuations in the
nonlinear regime. Many authors have addressed this issue using the Spherical
Collapse Model, e.g.,
\cite{Mota:2004pa,Nunes:2004wn,Abramo:2007iu,Creminelli2010,Basse2011}. 
Independently of the approach used, the most important point we have to
observe is when DE fluctuations actually become nonlinear, i.e., $\delta_e
\gtrsim 1$. The key property that controls this behaviour is the
effective sound speed, $c_{{\rm eff}}$, which was shown can be constrained
using future data on the abundance of galaxy clusters
\cite{Abramo2009a,Basse2012}. 

As we showed in the linear regime, DE perturbations in models with
non-negligible $c_{{\rm eff}}$ are at least
few orders of magnitude smaller than $\delta_{\rm m}$ on small scales. We also
verified that, in this case, neglecting DE perturbations changes the value of
the linearly evolved $\delta_{\rm m}$ only about $0.02\%$.  Moreover its value
is proportional to the value of $\phi$, which, before DE domination, is known to
remain approximately constant even during the nonlinear regime
\cite{Brainerd:1993kf,Bagla:1993ee}. Hence we expect that the contribution of
$\delta_{\rm e}$ for the gravitational potential will be very small during the
nonlinear regime. For this reason, in the case of $c_{{\rm eff}}=1$, we assume
that DE perturbations can be neglected for the nonlinear evolution. 

For the case of $c_{{\rm eff}}=0$, DE perturbations can have the same
order of magnitude of matter perturbations, hence they can not be neglected,
otherwise the linearly evolved $\delta_{\rm m}$ can change up to $20\%$.
Evidently an intermediate behaviour should appear when $c_{{\rm eff}}\gtrsim
c_{\rm nl}$, in this case one could incorporate the linear evolution DE
perturbations with the nonlinear evolution of DM \cite{Basse2011}. Here we want
to focus on the two limiting cases: the one where $\delta_{\rm e}$ is most
unimportant for the nonlinear evolution of DM, $c_{{\rm eff}}=1$, and the other
where $\delta_{\rm e}$ is the most important for it.

We use the fluid approach to solve the SC model derived from the
Pseudo-Newtonian Cosmology (PNC) \cite{Abramo:2007iu,Abramo:2008ip}, which is
suitable to treat the nonlinear evolution of perfect fluids with relativistic
pressure on small scales. The equations for a system with matter and DE
fluctuations, with $c_{{\rm eff}}=0$, can be written as:
\begin{equation}
\dot{\delta}_{\rm m}+\theta\left(1+\delta_{\rm m}\right)=0\,,
\label{mat_cont}
\end{equation}
\begin{equation}
\dot{\delta}_{\rm e}-3\mathcal{H}w\delta_{\rm e}+\theta\left(1+w+\delta_{\rm
e}\right)=0\,,
\label{de_cont}
\end{equation}
\begin{equation}
\dot{\theta}+\mathcal{H}\theta+\frac{\theta^{2}}{3}=-\frac{3\mathcal{H}^{2}}{2}
\left(\Omega_{\rm m}\delta_{\rm m}+\Omega_{\rm e}\delta_{\rm e}\right)\,.
\label{euler}
\end{equation}
For the case of negligible DE perturbations we only need to evolve
Eqs.~\eqref{mat_cont} and \eqref{euler}, setting $\Omega_{\rm e}\delta_{\rm
e}=0$ in the latter, in this case we have the usual SC model for dark matter
\cite{Gunn:1972sv,Padmanabhan,Percival:2005vm}.

The most relevant quantity we compute using the SC model is the critical density
contrast, $\delta_{\rm c}$, defined by
\begin{equation}
\delta_{\rm c}\left(z_{\rm c}\right)=\delta_{\rm mL}\left(z_{\rm c}\right)\,,
\end{equation}
where $\delta_{\rm mL}$ is the linearly evolved matter density contrast with
initial conditions such that non-linearly evolved $\delta_{\rm m}$ has vertical
asymptote at $z_{\rm c}$, i.e., 
\begin{equation}
\underset{z\rightarrow z_{\rm c}^{+}}{\lim}\delta_{\rm
m}\left(z\right)=\infty\,.
\end{equation}
Numerically we need to determine some value above which we consider $\delta_{\rm
m}$ has diverged; we observed that choosing $10^{5}$ gives the classical value
in Einstein-de-Sitter universe, $\delta_{\rm c}\simeq1.686$, with an error
of at most $0.5\%$ for $0\le z\le4$.

The function $\delta_{\rm c}$ depends both on the linear and nonlinear
evolution. Although we do not have a framework to
solve the relativistic SC model, we can compare the PNC and GR predictions for
the linear evolution. It is clear that PNC will not give accurate results for
scales of order of magnitude comparable with the horizon, however, for
$k=0.1\,\mbox{h}\,\mbox{Mpc}^{-1}$, the values of $\delta_{\rm m}$ given by PNC
differ from GR at most $0.1\%$ and decreases for smaller scales. Hence computing
$\delta_{\rm c}$ using Eqs.~\eqref{mat_cont}$-$\eqref{euler} and its
linearised version should be a good approximation.

\begin{figure}[tbp]
\centering
\includegraphics[scale=0.85]{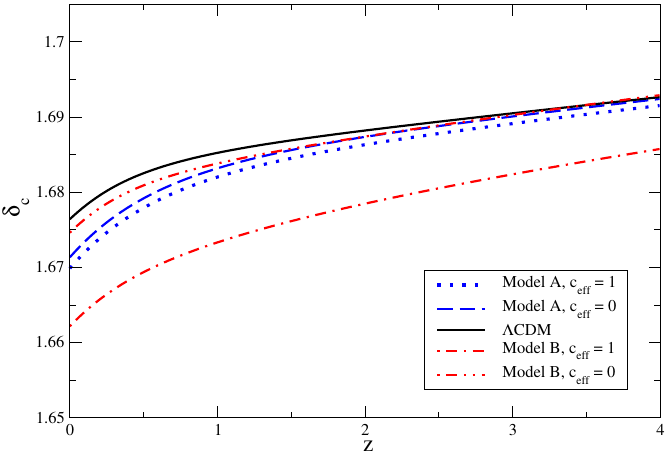}
\caption{Evolution of $\delta_{\rm c}\left(z_{\rm c}\right)$ for the different
models. Line styles and colour are shown in the legend box.
}
\label{fig:delta_crit}
\end{figure}

Now we need to provide the initial conditions to evolve the nonlinear equations.
We choose values of $\delta_{\rm m}\left(a_{\rm i}\right)$ in order to get $0\le
z_{c}\le4$. As we discussed in section 3.2, for the case of null
$c_{{\rm eff}}$ we can determine $\delta_{\rm e}\left(a_{\rm i}\right)$ using
Eq.~\eqref{de_dm_old_sol},  use
$\theta\left(a_{\rm i}\right)=-\dot{\delta}_{\rm m}\left(a_{\rm i}\right)$,
assuming        $\delta_{\rm m}\propto a$.
For the case of non-negligible $c_{{\rm eff}}$ we assume $\delta_{\rm m}\propto
a^p$ with $p$ determined by Eq.~\eqref{p_c1}. It is important to observe
that all these relations can be derived from the PNC equations, 
Eqs.~\eqref{mat_cont}$-$\eqref{euler}.

We show the critical density contrast in function of $z$ in
Fig.~\ref{fig:delta_crit}. For the cases of $c_{{\rm eff}}=0$ DE perturbations
clearly make $\delta_{\rm c}$ closer to the $\Lambda$CDM values. This happens
because, while at background level EDE changes the expansion rate and lowers
$\Omega_{\rm m}$, DE perturbations contribute to the gravitational potential,
compensating the change in background. Note that the values of $\delta_{\rm c}$
that we found are consistent with small departures (less than $1\%$) from
$\Lambda$CDM values \cite{Francis:2008ka,Pace2010}, even in the presence of
large EDE fluctuations.

\subsection*{The total mass of halos}

In models with clustering DE we must care about its contribution to the total
mass of the halo. The presence of DE can either add or subtract mass from the
forming DM halo. For the case of negligible $c_{{\rm eff}}$,
Ref.~\cite{Creminelli2010} argues that DE contribution is constant in time and
evaluates it at the virialization radius, $R_{\rm vir}$.

The computation of $R_{\rm vir}$ depends on the details of the virialization
process. In the usual SC model, in an EdS universe, the virialization radius is
half of the turnaround radius, $R_{\rm vir}=R_{\rm ta}/2$. Once the
virialization is reached one can define the DM overdensity of the formed halo.
There are two common definitions: $\Delta_{\rm
V}(z_{\rm c})= (\rho _{m}(z_{\rm v})+\delta \rho_{\rm m}(z_{\rm v}))/\rho
_{m}(z_{\rm c})\simeq178$, where
$z_{\rm v}$ is the redshift of virialization, and $\Delta_{\rm V}(z_{\rm
v})=(\rho _{m}(z_{\rm
v})+\delta \rho_{\rm m}(z_{\rm v}))/\rho _{m}(z_{\rm v})\simeq147$, see
Refs.~\cite{Lee2010b,Meyer:2012nw} for details and discussion.
However when DE is present these values change in time and the virialization
process also depends on the properties of DE
\cite{Lahav:1991wc,Maor:2005hq,Creminelli2010,Basse2012}.

Here we will use the usual relation between turn-around and virialization radius
$R_{\rm vir}=R_{\rm ta}/2$. Although this is not valid in non-EdS models and the
values of $\Delta_{\rm V}$ can be considerably different, this simplified
relation gives a good approximation for the fraction of DE mass to the DM mass:
\begin{equation}
\epsilon=\frac{M_{\rm e}}{M_{\rm m}}\,.
\label{eqn:eps}
\end{equation}
For instance, we checked that changing $R_{\rm vir}$ by $5\%$, which is common
for inhomogeneous DE models
Ref.~\cite{Basse2011}, changes $\epsilon$ by $1\%$.

After solving the system of Eqs.~\eqref{mat_cont}$-$\eqref{euler} we find the
value $R$ using the conservation of DM mass:
\begin{equation}
\delta_{\rm m}+1=(\delta_{\rm m_{\rm i}}+1)\left(\frac{a}{a_{\rm i}}
\frac{R_{\rm i}}{R}
\right)^3\;.
\end{equation}
The virialized mass associated with DM is then given by:
\begin{equation}
M_{\rm m}=4\pi \int _0 ^{R_{vir}} dR R^2 (\rho_{\rm m}+\delta\rho_{\rm m})\,.
\end{equation}
Now we also have to define the DE mass contained in the same region where the DM
halo has been formed, $M_{\rm e}$. In Ref.~\cite{Creminelli2010} it is defined
as the matter associated only with the fluctuations of DE, which we will call
\begin{equation}
M_{\rm e_{\rm P}}= 4 \pi \int _0 ^{R_{vir}} dR R^2 \delta \rho_{\rm e}\,.
\label{de_mass_pert}
\end{equation}
However, this choice makes a discrimination between the DE and DM contributions.
The latter is computed using its total
density, $\rho_{\rm m} + \delta \rho_{\rm m}$, whereas the former only considers
the fluctuation contribution.  Treating the two fluids on equal foot, the total
contribution of DE mass can be evaluated based on the
so-called active density in the Poisson equation in the presence of relativistic
pressure, $\nabla^2 \phi=4\pi G(\rho
+ 3p)$. Hence we have:
\begin{equation}
M_{\rm e_{\rm T}}= 4 \pi \int _0 ^{R_{vir}} dR R^2
\left[ (1+3w)\rho_{\rm e}+(1+3c^2_{{\rm eff}})\delta \rho_{\rm e}\right]
\label{de_mass_tot}
\end{equation}
The motivation to consider the total mass (background plus perturbative
contributions) is that the astrophysical processes associated with galaxy
clusters are sensitive to the total gravitational potential. This
interpretation is also carried out by Ref.~\cite{Chernin2009} in order to
estimate the mass of the Local Group. Although this background contribution may
be debatable we decided to evaluate it in order to verify whether it can
produce a non-negligible effect. For models with  $c_{{\rm eff}}=1$,
DE perturbations are much smaller than unity at the virialization time, hence
they can be neglected in Eq.~{\eqref{de_mass_tot}}.

\begin{figure}[tbp]
\centering
\includegraphics[scale=0.65]{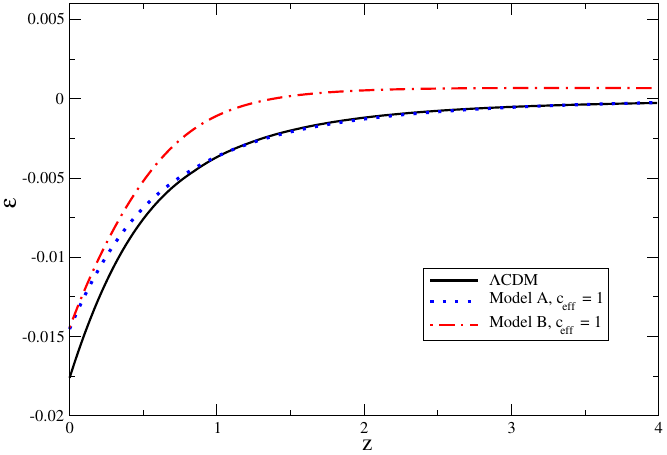}
\includegraphics[scale=0.65]{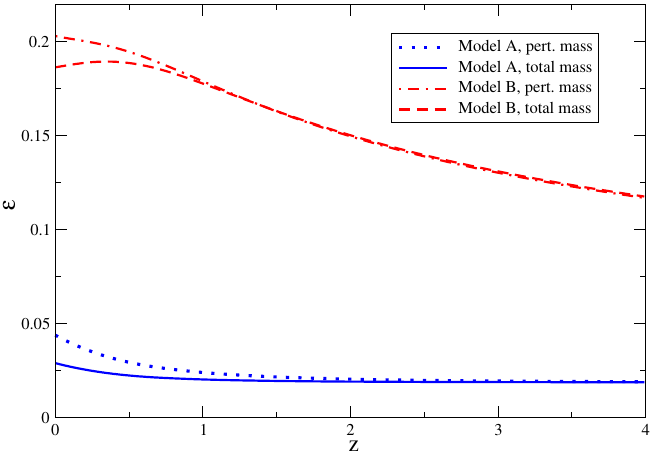}
\caption{Evolution of $\epsilon(z)$. Left panel: $\Lambda$CDM model and models A
and B with $c_{\rm eff}=1$. Right panel: Model A only with perturbation
contribution (dotted blue line) and total contribution (solid blue line),
Model B only with perturbation contribution (dot-dashed red line) and with
total contribution (dashed red line), all with $c_{\rm eff}=0$. 
}
\label{fig:epsilon}
\end{figure}

In Fig.~\ref{fig:epsilon} we show the evolution of the quantity $\epsilon$ for
models with homogeneous DE (left panel), in which case we compute $\epsilon$
using Eq.~{\eqref{de_mass_tot}}, and for models with inhomogeneous DE,
computing $\epsilon$ both with Eq.~{\eqref{de_mass_pert}} and
Eq.~{\eqref{de_mass_tot}} (right panel). Note that due to
the background contribution in Eq.~{\eqref{de_mass_tot}}, even in the case of
$c_{{\rm eff}}=1$ and $\Lambda$CDM models, DE can subtract only about $1\%$ of
DM mass. However this effect appears only at low redshifts, when $\Omega_{\rm
e}\sim \Omega_{\rm m}$. For the case of $c_{{\rm eff}}=0$, the fraction of DE in
a DM halo can be much larger, but now adding mass to the halo. In these models
the DE contribution does not go to zero at higher $z$ because of the combination
of two factors: while $\Omega_{\rm e}$ decreases, but not to negligible values,
$w$ decreases as well, which in turn, according to Eq.~{\eqref{de_dm_old_sol}},
makes the ratio $\delta_{\rm e}/\delta_{\rm m}$ grow.

This is a characteristic behaviour of EDE models. As already shown in
Ref.~\cite{Creminelli2010}, for the perturbative contribution and constant $w$,
around $-1$, $\epsilon$ is at most $0.05$ and tends to zero as $z$ grows. In
Ref.~\cite{Basse2012}, for models with $w=-0.8$ and $c_{{\rm eff}}^{2}=10^{-6}$,
the authors claim that maximal value of DE mass fraction in halos is $0.1\%$.
However they assume DE fluctuations are small and can be treated within linear
theory, which is not valid for the models with $c_{{\rm eff}}=0$ we are
studying.

\section{Abundance of halos}\label{sect:mf}
In this section we study the effect DE fluctuations on the abundance of halos.
We decided to parametrize the mass function using the Sheth \& Tormen
(ST) prescription \cite{Sheth1999,Sheth2002,Sheth2001}: 
\begin{equation}
\frac{dn}{dM}=-\sqrt{\frac{2a}{\pi}}A
\left[
1+
\left(
\frac{a\delta _c^2}{D^2\sigma^2_M}
\right)^{-p}
\right]
\frac{\bar{\rho}_m}{M^2}\frac{\delta _c}{\sigma_M}
\frac{d\ln \sigma_M}{d \ln M}\exp
\left(
\frac{-a\delta _c^2}{2D^2\sigma^2_M}
\right)\,
\end{equation}
where $a=0.707$, $p=0.3$, $A=0.2162$, 
\begin{equation}
\sigma^2_M = \frac{1}{2\pi^2}\int_0^{\infty}dkk^2 W^2\left(kR\right)P(k)
\end{equation}
is the squared variance of the matter power spectrum, P(k), which we computed
using the BBKS transfer function \cite{Bardeen:1985tr}, smoothed with a top-hat
window function, $W(kR)=3(kR)^{-3}(\sin(kR)-kR \cos(kR))$, where $R$ is the
scale enclosing the mass $M=(4\pi/3) R^3\bar{\rho}_m$ and $\bar{\rho}_{\rm m}$
is the
comoving matter density. The ST mass function depends critically on the
linear overdensity parameter $\delta_{\rm c}$ of dark matter, therefore also a
small variation of this quantity will give a huge effect on the high-mass tail
of the mass function. 

In order to clearly observe the effects of DE fluctuations we compute the number
density of objects above a given mass at fixed redshift:
\begin{equation}
n(>M)=\int_M^{\infty} \frac{dn}{dM'}dM'\;.
\end{equation}
We choose four different redshifts, namely $z=0, 0.5, 1, 2$. The actual number
of halos also depend on a integral over the comoving volume,
hence it depends both on background and perturbative properties of DE. For the
models we consider the comoving volume is always smaller than in $\Lambda$CDM,
see Fig.~\ref{fig:volumes}.

We  adopt as reference model the $\Lambda$CDM model with normalisation of the
matter power spectrum $\sigma_8=0.776$, in agreement with CMB measurements by
the WMAP team \cite{Komatsu2011,Larson2011}. Since the background history and
therefore the growth factor for the EDE models differ from the $\Lambda$CDM
models, perturbations will evolve differently in the different classes of
models. Therefore we decide to adopt the CMB normalisation: we fix the same
amplitude of the perturbations at the CMB epoch and we rescale it by the ratio
of the different growth factors. More quantitatively we have
\begin{equation}
\sigma_{8,\rm DE}=\sigma_{8,\Lambda\rm CDM}\frac{D_{\Lambda\rm CDM}(z_{\rm
dec})}{D_{\rm DE}(z_{\rm dec})}\;.
\end{equation}
In Tab.~\ref{tab:sigma8} we show the normalisation of the power spectrum for the
dark energy models considered in this work.

\begin{table}
\caption{Table of the normalisation of the matter power spectrum
for the different models analysed.}
 \label{tab:sigma8}
 \begin{center}
  \begin{tabular}{c|c}
   \hline
   \hline
   Model & $\sigma_8$\\
   \hline
   A, $c_{\rm eff}=0$ & 0.761 \\
   A, $c_{\rm eff}=1$ & 0.698 \\
   B, $c_{\rm eff}=0$ & 0.674 \\
   B, $c_{\rm eff}=1$ & 0.532 \\
   \hline
   \hline
  \end{tabular}
 \end{center}
\end{table}

\begin{figure}[tbp]
 \centering{}
 \includegraphics[scale=0.63]{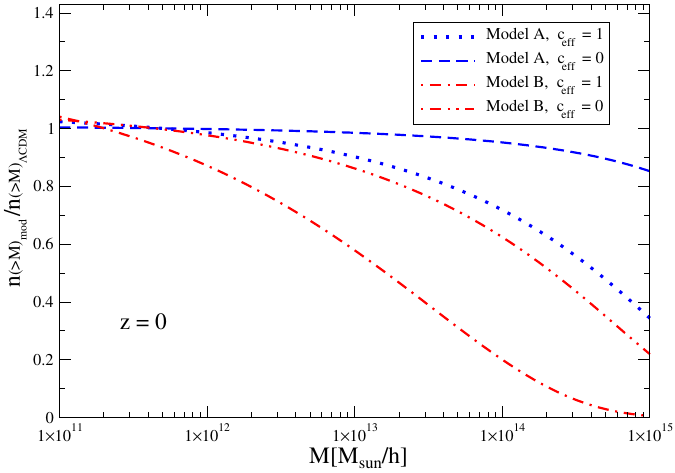}
 \includegraphics[scale=0.63]{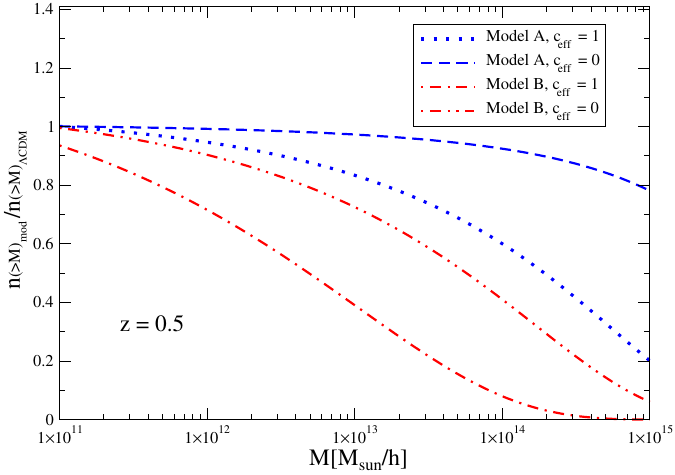}
 \includegraphics[scale=0.63]{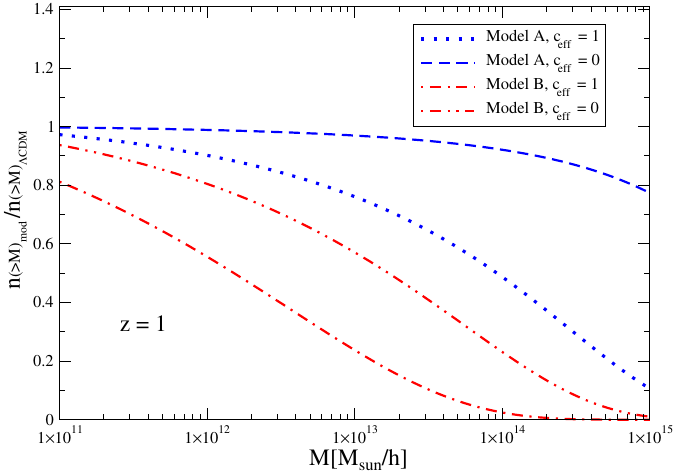}
 \includegraphics[scale=0.63]{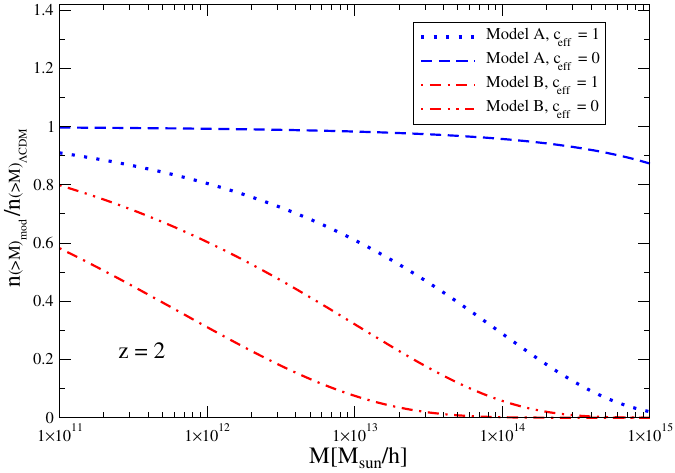}
 \caption{Number density ratio between the EDE models and the
reference $\Lambda$CDM model. The red dashed (blue short-dashed) curve shows
results for Model A with $c_{\rm eff}=0$ ($c_{\rm eff}=1$), while the cyan
dotted (orange dot-dashed) curve represents Model B for $c_{\rm eff}=0$ ($c_{\rm
eff}=1$). The upper left (right) panel shows results for $z=0$ ($z=0.5$), while
the lower left (right) panel shows results for $z=1$ ($z=2$).}
\label{fig:mf}
\end{figure}

As one can notice, the Model B is the one differing most from the $\Lambda$CDM
model. The matter power spectrum normalisation when $c_{\rm eff}=1$ is 30\%
lower than the reference, therefore we expect that the mass function will be
significantly different from the reference $\Lambda$CDM model. In
Fig.~\ref{fig:mf} we show the ratio between the number density of dark matter
halos for the EDE models and the $\Lambda$CDM model. We refer to the caption for
the different line-styles and colours.

It is immediately clear that for Model B structures are strongly suppressed at
all redshifts with respect to the $\Lambda$CDM model. Model B shows a lack of
objects already at galactic scales ($M\simeq 10^{11}-10^{12}~M_{\odot}/h$)
while at cluster scales ($M~\approx 10^{15}~M_{\odot}/h$) the case with $c_{\rm
eff}=1$ has basically no objects (see
the upper left panel). At higher redshifts, the number of objects decreases
considerably also at very low scales (see lower right panel for the results at
$z=2$).

Model A suppresses structures less than Model B and 
for low mass objects at $z=0$ the Model A with $c_{\rm eff}=1$ is quite
similar to the Model B with $c_{\rm eff}=0$, but for masses of the order of
$10^{15}~M_{\odot}/h$ they differ by about $30\%$. With the increase of the
redshift, differences between the two models of EDE tend to increase.
The most similar model to the $\Lambda$CDM one is Model A with $c_{\rm eff}=0$.
Also at very high masses differences are at most about $20\%$. It is also
important to bear in mind that the actual number of halos will also depend on
the comoving volume, Fig.~\ref{fig:volumes}, which for EDE models is always
smaller than in
$\Lambda$CDM models. For low redshifts, $z=0$ and $z=0.5$, the differences in
the comoving volume is very small, but for $z=1$ and $z=2$ the actual number of
halos in EDE models will be even smaller than the ratios in
the lower panels of Fig.~\ref{fig:mf} suggest.

It is also worth to note that models with $c_{\rm eff}=0$ have more objects than
the case with $c_{\rm eff}=1$. This is easily explained in terms of the growth
factor and of the evolution parameter $\delta_{\rm c}$. Again the
general effect of large DE fluctuations is to compensate the change in the
background evolution by enhancing the gravitational attraction.

\subsection*{The contribution of DE mass to the mass functions} 

As already observed in Ref. \cite{Creminelli2010}, if DE can cluster it
also contributes to the halo mass, thus we must compute the correction to the
mass function due to this extra component. The actual contribution of DE
crucially depends on whether it virializes and the time scale of this process.
Moreover the merging history of halos formed at different times
\cite{Lacey:1993iv}, which could contain different amounts of DE, should also be
taken into account. A complete and accurate description of this corrections is a
complex task, which depends on nature of DE, and is beyond the scope of this
paper. Here we will compute a straightforward correction for a scenario where,
in the case of $c_{\rm eff}=0$, we assume that DE virializes together with DM on
the same time scale. Hence, once the halo is formed, DE contribution is assumed
to remain constant. This is probably the case in which DE fluctuations will
mostly influence structure formation and consequently the mass function.  

Given a halo with DM mass $M$, its total mass will be
$M(1+\epsilon(z))$, with $\epsilon(z)$ defined in Eq.~\eqref{eqn:eps}.
Therefore, given the original DM mass function, $\frac{dn}{d\ln{M}}$, we assume
the corrected mass function is given by:
\begin{equation}
\frac{dn_c}{d\ln{M}} (z,M)= 
\frac{dn}{d\ln{M}}  \left(z,M(1-\epsilon)\right)\;. 
\label{eqn:MFtot}
\end{equation}
We call the attention of the reader to the minus sign in $M(1-\epsilon)$.
Although the mass of a halo is changed by $M \rightarrow M(1+\epsilon)$ the use
of this mass redefinition in the mass function would produce wrong results. For
a positive $\epsilon$ the halos become more massive than predicted by a model
in which only DM clusters, hence more massive halos are expected.
However, if one redefines the mass function using $M(1+\epsilon)$, fewer massive
halos would be predicted, which is just the opposite of what is expected.
Therefore the natural correction of the mass function is the change of variable
$M \rightarrow M(1-\epsilon)$.

\begin{figure}[tbp]
 \centering{}
 \includegraphics[scale=0.65]{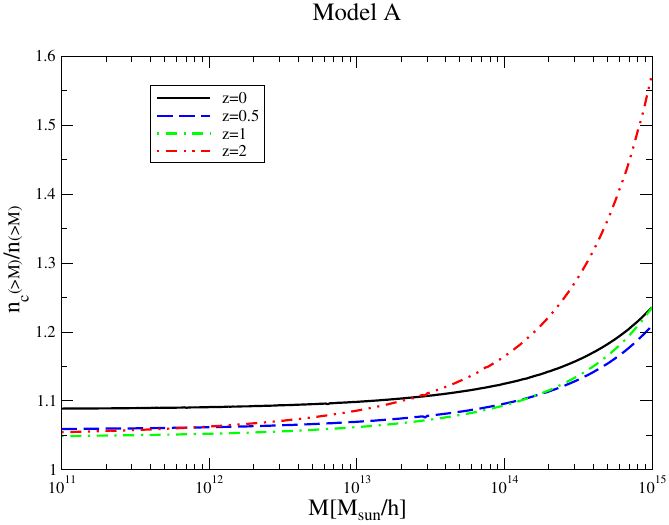}
 \includegraphics[scale=0.65]{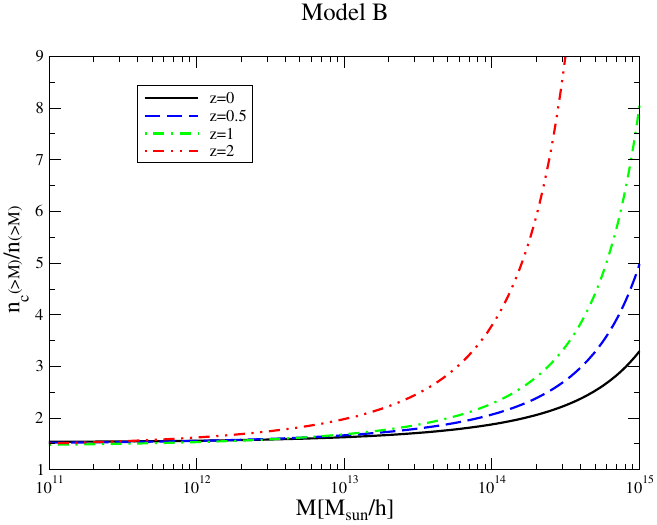}
  \includegraphics[scale=0.65]{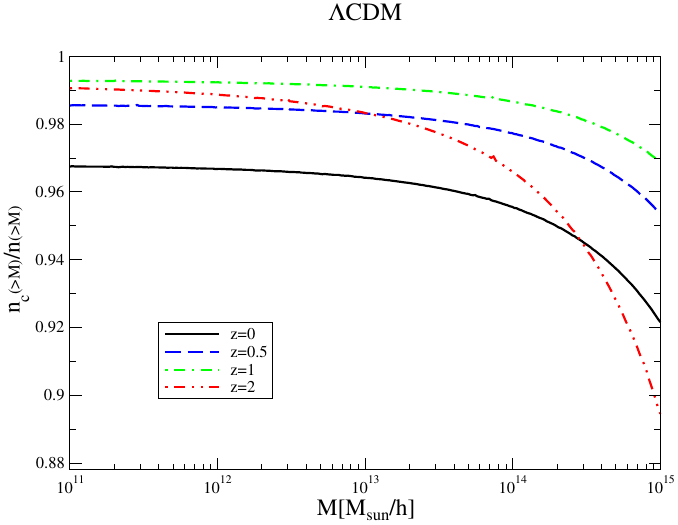}
 \caption{Ratio of comoving number density above mass $M$ at fixed redshifts,
computed with the mass function correction, Eq.~\eqref{eqn:MFtot}, $n_c(>M)$ to
the comoving number density in the same model without the correction $n(>M)$. 
In Model B the corrections reach $60$ times at $z=2$.} 
\label{fig:MFcorr}
\end{figure}

In Fig.~\ref{fig:MFcorr} we show how much is the correction in the  comoving
density of objects above mass $M$ relative to the same model without the
correction, $n_c(>M)/n(>M)$.  We present the results for models A and B, with
$c_{\rm eff}=0$, only for the perturbative contribution (top panels).
Although it is not clear whether the background energy
density of DE is a stable contribution to the total mass of the halo, i.e., that
is constant through the history of the halo, in order to have an idea of
its influence on the abundance of halos, we  also show the corrections due
to the background contribution in the $\Lambda$CDM model (bottom panel).  

As can be seen in the these three cases, it is clear that even a small DE
contribution for the halo mass can produce drastic changes at high mass. For
Model B, the one with largest $\epsilon$, $n_c$ can be many times larger then
its version without the mass correction. In Model A the corrections are below
$10\%$ for small masses but can reach almost $60\%$ for high masses at high $z$.
For $\Lambda$CDM, the corrections can be about $3\%$ at $z=0$ and low masses,
reaching more than $10\%$ for very massive halos at high $z$. 

It is also clear that, although $\epsilon$ gets smaller at higher $z$, the
corrections do not necessarily diminish, as can be seen for the three cases with
$z=2$. Since the growth function, $D$, gets smaller for high $z$, the
exponential
decay of the mass functions is shifted to lower mass regions, hence even a small
value of $\epsilon$ at high $z$ produces large modifications. The comprehension
of this effect is very important because, despite the fact
that very few massive galaxy clusters are expected at high $z$, the detection of
a single massive distant galaxy cluster can be used to rule out DE models
\cite{Mortonson:2010mj}. 

\begin{figure}[tbp]
 \centering{}
 \includegraphics[scale=0.65]{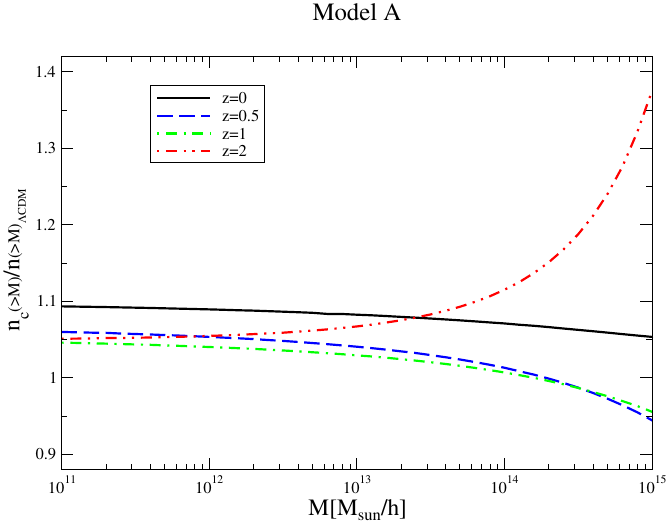}
 \includegraphics[scale=0.65]{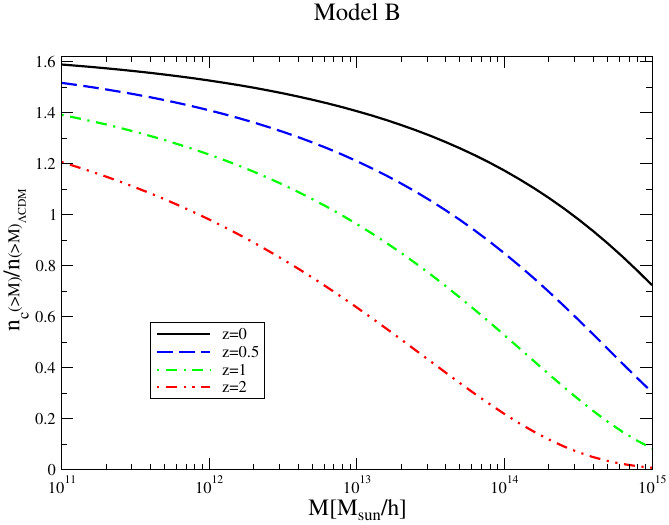}
 \caption{Ratio of comoving number density above mass $M$, computed with the
mass function correction Eq.~\eqref{eqn:MFtot}, $n_c(>M)$ to the comoving number
density in the $\Lambda$CDM model.}
\label{fig:MFnew}
\end{figure}

Now that we observed that the corrections in mass functions can be important,
we have to ask ourselves whether they can change the behaviour that we
previously found without taking into account such corrections, i.e., that all
EDE models we consider have smaller density of objects then $\Lambda$CDM,
Fig.~\ref{fig:mf}. In Fig.~\ref{fig:MFnew} we plot the ratio of the corrected
values of number density, $n_c$, to the $n$ in $\Lambda$CDM model. For small
masses, the corrections indicate that our inhomogeneous EDE models actually
present more objects than $\Lambda$CDM. However, $n_c(>M)/n(>M)_{\Lambda CDM}$ 
decreases with mass, except for Model A at $z=2$. For Model B the suppression of
structures due to its low $\sigma_8$ value is so strong that, even with the
large corrections caused by its large values of $\epsilon$, at high masses it
always has fewer objects density than $\Lambda$CDM. 

In Model A a quite interesting behaviour can be observed. Since its $\sigma_8$
and growth function are not very different than in $\Lambda$CDM, the correction
due to DE mass turns out to be more important at high $z$ and rapidly increases
with mass. At $z=2$ the number density in Model A is about  $36\%$ larger than
in $\Lambda$CDM. Note that this effect occurs at high $z$ and is strongly mass
dependent. Ref.~\cite{Mortonson:2010mj} points out that if unexpected massive
clusters, within the $\Lambda$CDM or smooth quintessence paradigm, are present
only at high $z$, DE clustering may not provide a consistent description because
in these models cluster abundances are modified roughly by the same amount for
both low and high redshifts. However, as we just have observed, the corrections
on mass function due to DE mass may provide more abundant massive clusters at
high $z$, without drastic changes for low $z$. Hence, if in the future the
observation of a massive cluster falsifies $\Lambda$CDM and smooth quintessence
models, it still can interpreted as an evidence of clustering DE.

However, let us note once again that the actual number of objects also
depends on the comoving volume. Therefore the behaviour observed in
Fig.~\ref{fig:MFnew} can be modified, especially at high $z$. In order to
verify what is the observable effect, we finally compute the total number of
clusters above mass $M$ and redshift $z$:
\begin{equation}
N\left(>z,>M\right) = \int_z^{\infty}
dz'\frac{dV^2}{dz'd\Omega}\int_M^{\infty}dM'\frac{dn}{dM'}\,.
\label{n_tot}
\end{equation}

\begin{figure}[tbp]
 \centering{}
 \includegraphics[scale=0.65]{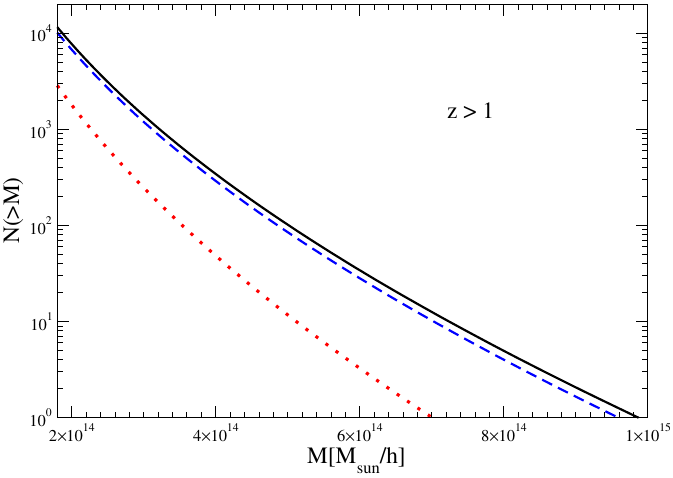}
 \includegraphics[scale=0.65]{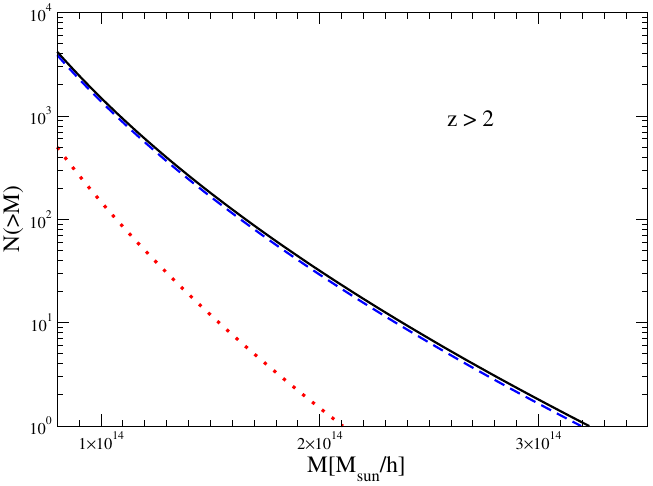}  
 \caption{Total number of halos above mass $M$ and $z=1$ (left panel) and $z=2$
(right panel), Eq.~\eqref{n_tot}. The solid black lines refers to $\Lambda$CDM
model, the blue dashed (red dotted) lines to Model A (Model B), with $c_{\rm
eff}=0$ and correct mass function due to the perturbative contribution only.}
\label{fig:n_tot}
\end{figure}

In Fig.~\ref{fig:n_tot} we show  the total number of halos above mass $M$ and
$z=1$ (left panel) and $z=2$ (right panel) for $\Lambda$CDM, Model A and Model
B. The values for Model A and Model B are computed using the corrected mass
function for the perturbative contribution only. For $z>1$ the EDE models
present fewer objects than $\Lambda$CDM. The same is true for $z>2$, however
Model A and $\Lambda$CDM are much more similar. Therefore, in these
specific examples, we can see that the correction in the mass function due to
DE mass is not enough to produce more massive and distant clusters than
$\Lambda$CDM. Anyhow the main lesson we should take from these results is that
if DE possesses fluctuations, its contribution for the halo mass can
substantially change the abundance of massive galaxy clusters. The proper
description of this effect is of major importance when testing inhomogeneous DE
models with observations of such objects. 

\section{Conclusions}\label{sect:conc}

In this paper we have studied the influence of inhomogeneous EDE both in linear
and nonlinear stages of structure formation. We have evaluated the matter
growth, the ISW effect, the contribution of DE fluctuations for the total mass
of the halos, the halo abundance relative to $\Lambda$CDM model and its
corrections due the extra DE mass contribution. 

We have showed that the presence of large EDE fluctuations, i.e., models with
$c_{\rm eff}=0$, have the general property of making DM growth, ISW effect and
halo abundance more similar to the predictions of the $\Lambda$CDM model than
homogeneous EDE models, those with $c_{\rm eff}=1$. In Model A, the one more
similar to $\Lambda$CDM in the background evolution, the differences in $f=d\ln
D/d\ln a$ against $\Lambda$CDM predictions are below $5\%$ level, and the impact
of DE fluctuations is even smaller. Eventually surveys like Euclid, which can
provide data on $f$ with $1\%$ precision \cite{Amendola:2012ys}, will be able to
distinguish between homogeneous and inhomogeneous EDE models.

The analysis of the number density of halos initially showed that all
EDE models provide fewer density of massive clusters, independently of the
redshift considered. For models with nearly smooth EDE this conclusion remains
valid. However, for inhomogeneous EDE models, once we account for the extra halo
mass associated with DE fluctuations, which we showed that can be of the order
of $10\%$, and make the corresponding correction in the mass function, this
situation may change. We saw that, at high redshifts, the corrected number
density can be larger than in $\Lambda$CDM. This was observed for our Model A
with $c_{\rm eff}=0$ at $z=2$. However, after computing the total number of
halos, the impact of this correction is supplanted by the effect of the smaller
comoving volume of EDE models and $\Lambda$CDM still presents more massive
objects. 

It is important to stress that, in principle, the magnitude of DE fluctuations
that we found and the corresponding impact on observables that we have studied is
specific of EDE models. If DE has non-negligible energy density at intermediate
and high redshifts the equation-of-state parameter $w$ is close to $0$, which,
according with Eq.~\eqref{de_dm_old_sol}, enhances the magnitude of
DE fluctuations. However we also showed that inhomogeneous EDE models actually
make predictions more similar to $\Lambda$CDM than their homogeneous
counterparts. Therefore we conclude that if the accelerated expansion is caused
by an inhomogeneous EDE model it will be challenging to distinguish it from the
Cosmological Constant.

\section*{Acknowledgements}
RCB thanks CNPq and FAPERN for the financial support. RCB is also grateful to
the Institut f\"ur Theoretische Astrophysik of the Heidelberg University 
for the warm hospitality and support during a visit when this work was
initiated. FP is supported by STFC grant ST/H002774/1.

\bibliographystyle{JHEP}
\bibliography{referencias}

\end{document}